\documentclass[preprint,12pt,3p]{elsarticle}

\usepackage{amsmath}
\usepackage{amssymb}
\usepackage{graphicx}
\usepackage{xcolor}
\usepackage{caption}
\usepackage{wrapfig}
\usepackage{subcaption}
\usepackage{macros_vlastni}
\usepackage{amsfonts}
\usepackage{ulem}
\usepackage{appendix}

\newcommand\bmi[1]{\textbf{\textit{#1}}}

\newcommand\rhs{{r.h.s.{~}}}

\newcommand\wrt{{w.r.t.{~}}}

\newcommand\ie{{\it{i.e.~}}}
\newcommand\eg{{\it{e.g.~}}}
\newcommand\cf{{{cf.~}}}

\newcommand\ext{{\rm{ext}}}
\newcommand\eqe{{\rm{eq},\veps}}

\newcommand{\ol}[1]{\overline{#1}}

\def\ub{{\bmi{u}}}
\def\vb{{\bmi{v}}}
\def\wb{{\bmi{w}}}
\def\fb{{\bmi{f}}}

\def\nb{{\bmi{n}}}

\def\eb{{\bmi{e}}}

\def\qb{{\bmi{q}}}

\def\Ab{{\bmi{A}}}

\def\Eb{{\bmi{E}}}

\def\Wb{{\bmi{W}}}

\def\Ib{{\bmi{I}}}

\def\Eb{{\bmi{E}}}

\def\Lb{{\bmi{L}}}

\def\Hdb{{\bf{H}}^1}
\def\Hop{{{\rm I} \kern-0.2em{\rm H}}}%
\def\Hpdb{{\bf{H}}_\#^1}

\def\Mcal{\mathcal{M}}
\def\Tcal{\mathcal{T}}

\def\intY{\:\sim \kern-1.17em \int}
\def\intYs{- \kern-0.82em \int }
\def\intYsmall{{\small \sim} \kern-0.95em \int}
\newcommand\Tuftxt{\Tcal_\veps\,}

\begin{document}

\title{Homogenization based two-scale modelling of ionic transport in fluid saturated deformable porous media}

\author[1]{Jana Turjanicov\'a\corref{cor1}}
\ead{turjani@ntis.zcu.cz}

\author[1]{Eduard Rohan}
\ead{rohan@kme.zcu.cz}

\author[1]{Vladim\'{\i}r Luke\v{s}}
\ead{vlukes@kme.zcu.cz}

\address[1]{European Centre of Excellence, NTIS -- New Technologies for
Information Society, Faculty of Applied Sciences, University of West Bohemia,
Univerzitni\'{\i} 8, 30614 Pilsen, Czech Republic}

\cortext[cor1]{Corresponding author}

\begin{abstract}
	The paper deals with the homogenization of deformable porous media saturated by two-component electrolytes. The model relevant to the microscopic scale describes steady states of the medium while reflecting essential physical phenomena, namely electrochemical interactions in a dilute Newtonian solvent under assumptions of a small external electrostatic field and slow flow. 
	The homogenization is applied to a linearized micromodel, whereby the thermodynamic equilibrium represents the reference state. Due to the dimensional analysis, scaling of the viscosity and electric permitivity is introduced, so that the limit model retains the characteristic length associated with the pore size and the electric double layer thickness. The homogenized model consists of two weakly coupled parts: the flow of the electrolyte can be solved in terms of a global pressure and streaming potentials of the two ions, independently of then the solid phase deformations which is computed afterwards for the fluid stress acting on pore walls. The two-scale model has been implemented in the \textit{Sfepy} finite element software. The numerical results show dependence of the homogenized coefficients on the microstructure porosity. By virtue of the corrector result of the homogenization, microscopic responses in a local representative cell can be reconstructed from the macroscopic solutions.
\end{abstract}


\maketitle
\section{Introduction}
\label{intro}

Modelling the transport of an electrolyte solution through a porous medium (TEPM) is a multiscale nonlinear problem with obvious multiphysics features. The problem is of interest for a wide range of science fields, including geosciences, environmental engineering, physiology and tissue biomechanics, material science and namely chemical engineering. Moreover, there are challenging industrial applications related to energy storage (batteries), extraction of renewable energy from salinity differences, or corrosion of reinforced concrete structures. The technology-related areas of engineering and scientific research require a quantitative analysis of the TEPM using computational modelling which allows to capture influences of microstructure related phenomena on the macroscopic properties and behavior at the macroscopic level. In this context, the homogenization of periodic, or locally periodic structures is one of the most relevant modelling approaches which lead to efficient computational algorithms. On one hand, the upscaling procedure enables to compute macroscopic tensorial coefficients respecting a given microstructure, on the other hand the downscaling procedure enables to interpret the macroscopic response at the microscopic level. For both these procedures, the so-called characteristic responses which are obtained as solutions of the representative volume elements (RVE) are needed.

The aim of this paper is to develop a two-scale computational model for the quasi-static transport of a two-component electrolyte solution trough a deformable porous medium, such that the upscaling and downscaling procedures allow for studying global and local effects in a response to the microstructure and material properties.

During the last decade, there appeared a significant body of literature devoted to the modelling of the TEPM. Here we comment only on those  publications which, as  we believe, are the most relevant and tightly related to the present work
\par As stated above, a remarkable part of the related research on the matter concerns the geosciences. One of the most recognized work in this field is the paper \cite{moyne2002electro} which relies on the homogenization procedure to derive a macroscopic model of expansive clays composed of a charged solid phase saturated by an electrolyte solution. This microscopic model includes equations describing electro-hydrodynamics coupled with the equation governing the flow of the electrolyte solution, ion electrodiffusion and electric potential distribution. Then the asymptotic homogenization is used to derive a two-scale model of electrokinetic phenomena, such as the electro-osmotic flow driven by the streaming potential gradient, the electrophoretic motion of mobile charges and the swelling induced by the osmosis. This model was later revisited in paper \cite{moyne2006two} with a more focus on the rigorous homogenization procedure and its analysis.  It should be noted that a similar problem, \ie the transport of an N-component electrolyte solution through porous rigid body subjected to a static electric field, was also studied in \cite{looker2006homogenization}, although no assumptions about the electric double layer were considered.

\par In biomechanical modelling, authors of \cite{lemaire2006multiscale} and \cite{lemaire2010multiphysical}, use a similar approach to study a bone fluid flows at two porosity levels in the cortical bone tissue. It is worth to note a possible application of that model in the studies of the  mechanosensing, \cf \cite{lemaire2010modelling}, and bone remodeling. These issues were treated in \cite{nguyen2009numerical}, where Biot's poroelastic theory applied to the three-dimensional anisotropic media in order to account for deformation induced fluid flows in osteonal matrix under the harmonic loading. 
Homogenization of the ionic exchange between the charged porous medium and the electrolyte solution was elaborated in \cite{lemaire2010modelling}, being motivated by mechanosensing.
A two-scale one-dimensional model for horizontal electro-osmotic flows in a number of thin horizontal slits was proposed in \cite{amirat2008electroosmosis}. Therein, the pressure gradient and a horizontal electrical field were recognized as the flow driving forces. Although this work is focused on one specific case and disregards any deformation of the solid part, it provides a useful insight to the homogenization of the electroosmotic law for different types of multi-component electrolytes.

\par Most of the works devoted to TEPM assume that an electrolyte saturates a rigid porous medium, see \cite{allaire2013asymptotic,allaire2013ion}, the deformation phenomenon or  evolving  porous structures were considered in a number of papers, see \eg \cite{ray2012multiscale,allaire2015ion}. Moreover, some other works, \eg \cite{andreasen2013topology,rohan-etal-CMAT2015,Sandstrom-Larssen-CMAME2016,Rohan-AMC} treating the fluid-structure interaction without any electro-osmotic, or electromechanical coupling established useful platforms for extensions of those particular models to account for the phenomena featuring the transport of electrolytes.
Such an extension was reported in \cite{allaire2015ion}, which is motivated by the study of nuclear waste disposal. In this work, the well known system of equations governing the ionic transport and extend it by elasticity of the solid part was introduced. The coupling between fluid motion and deformation of the solid matrix was also explored earlier in the work \cite{mikelic2012interface}. Therein the authors show that by a suitable choice of time scale, the deformation of porous medium becomes only weakly coupled to the electrokinetic system, which is advantageous for the model implementation and numerical simulations.

For completeness, let us note, that a non-stationary case model of ionic transport consisting of Stokes, Nerst-Planc and Poisson systems of equations was reported in papers \cite{ray2012rigorous, Schmuck-Bazant-SIAM2015} and \cite{frank2011numerical}, where the upscaling procedure was treated using the two-scale convergence. However, since the time response at multiple scales introduces further difficulties in the modelling, in our study, we account for steady state problems.

Although the ionic transport in porous structures is well-known problem, there are still some challenging issues deserving more attention. Most of the papers cited above concern the theoretical issues of the mathematical modelling, without numerical simulations.  On contrary, our interest lies in the implementation of a physically correct homogenized model, such that the upscaling and downscaling procedures are available for 3D microstructures. For this purpose we consider the model treated in \cite{allaire2015ion} and provide essential ingredients of the modelling, starting from the model definition at the microscopic level, pursuing the linearization which allows for using the two-scale homogenization.The derived the macroscopic model involves the effective medium parameters, such as poroelasticity coefficients, permeability, diffusivity and other coupling coefficients which satisfy the Onsager reciprocity relationships.


\par The paper is organized as follows. In  Section~\ref{sec:1} we provide a brief introduction in the physical phenomena and their mathematical descriptions which constitute the mathematical model of the two-component electrolyte solution transported in the deforming elastic skeleton. As the next step, in Section~\ref{sec:lin}, we introduce the dimensionless form of the mathematical model which is subject to the linearization procedure. The homogenization is reported in Section~\ref{sec_homog}; therein the principal results are presented, namely the local problems for the so-called corrector functions, formulae for computing the homogenized equations, and, finally, the macroscopic model.  In  Section~\ref{sec_numeric}, the effective coefficients relevant to both mathematical models are quantified for varying porosity the microstructure geometry. Furthermore, this section introduces the numerical solution of the macroscopic model and its recovery at the  microscopic level. Section~\ref{sec_conclusion} summarizes the results of this paper.  \ref{sec_A} is devoted to the unfolding operator. \ref{sec_B} clarifies the introduction of the scale parameter into the model.

\paragraph{Basic notations} Through the paper we shall adhere to the following notation. The position $x$ in the medium is specified through the coordinates
$(x_1,x_2,x_3)$ with respect to a Cartesian reference frame. We shall also use the microscopic (dilated)
Cartesian reference system of coordinates $(y_1,y_2,y_3)$. The gradients are employed, $\nabla_x = ( \pd / \pd x_i)$ and
$\nabla_y = (\pd / \pd y_i)$  alternatively.
As usually, the vectors will be  denoted by  bold letters, for instance,
$\ub(x)$ denotes the solid matrix displacement vector field depending on the
spatial variable $x$.  Moreover, the components of this vector will be denoted
by $u_i$ for $i=1,...,3$, thus $\ub = (u_i)$.  The Einstein summation
convention is used which stipulates implicitly that repeated indices are summed
over. By $\RR$ the real number set is denoted.
The differential volume and surface elements are
denoted by $\dV$ and $\dS$, respectively. 
Function spaces are
introduced subsequently in the text.
          
\section{Mathematical model}
\label{sec:1}
The porous medium occupies an open bounded domain $\Om \subset \RR^d$,  where $d=3$.
Without loss of generality, we may assume that the domain represent a  specimen shaped as a block $\Om=]0,L[^d$, which will enable us to impose periodic boundary conditions on $\pd\Om$.
According to the phases, $\Om$ splits into the fluid $\Om_f$ and solid $\Om_s$ parts, whereby both $\Om_s$ and $\Om_f$ are connected domains. By $\Gamma=\pd\Om_s\cap\pd\Omega_f$ we refer to the solid-fluid interface, and $\nb$ designates the unit  normal vector on $\Gamma$, being  outward to $\Omega_f$.  
The subscripts $\sqcup_s$ and $\sqcup_f$ will be used through the rest of the text also to denote the constants and variables belonging to the respective phases.

\subsection{Electrical double layer and the electrostatic potential of a phase}\label{potential-interface}
Due to the surface effects of charged skeleton $\Om_s$, in domain $\Om_f$ occupied by the electrolyte, the so-called electrical double layer (EDL) can be distinguished in the proximity of charged pore surfaces, were the ionic charge distribution is perturbed from the bulk. The EDL splits into two sub-layers, the Stern layer and the diffuse layer, The thickness of the EDL is related to the  
Debye parameter $\lambda_D$ which will be specified below. Within the EDL, the attraction is strong enough to influence particle movement. For a deeper physical insight we refer to \cite{hunter2001foundations}.


While in the solid phase the  electrostatic potential $\Psi_s$ is constant, the total electric potential $\Psi_f$ of the fluid phase is associated  with  the distribution of ions in the electrolyte.  Considering the effects of the EDL, the electrical potential $\Psi_f$ in the fluid varies strongly with the distance from to pore surface, since the ions show the tendency to arrange themselves to minimize their free energy. These effects, as illustrated in Fig.~\ref{fig_edl}, result in the Poisson-Boltzmann distribution of the electrical potential $\Psi_{\textrm{EDL}}$. 
In the bulk, \ie away from the solid-fluid interface, the electrical potential $\Psi_f = \Psi_{\textrm{bulk}}$ attains a constant value. 
{We can also introduce the electrostatic part $\Psi$ of the total potential in the fluid as the difference   $\Psi=\Psi_{\textrm{EDL}}-\Psi_{\textrm{bulk}}$.}

The fluid flow perturbs the electrostatic distribution of ions caused by the EDL and drags some ions in direction of the flow. This produces variations of the ionic concentrations along the flow direction, so that  the so-called streaming potential $\Phi_\vala$ can be defined for every $\vala$-th ionic species.


  If an external electrical field $\Ebex$ is present, another type of electrical potential is distinguished. By a construction, the so-called exterior (affine) potential $\Psiex = -x\cdot\Ebex $ can be introduced, such that $\Ebex=-\nabla\Psiex$. The imposed  field $\Ebex$ can usually be considered small, as compared with the fields in the EDL, so that a linearization  of the non-linear problem can be employed. 
We may summarize that the total potential $\Psi_f$ is given by the sum of electrostatic potential $\Psi$, streaming potentials $\Phi_\vala$, and exterior potential $\Psiex$, thus $\Psi_f=\Psi+\Phi_\vala+\Psiex$.


\subsection{Processes in the fluid phase}
\label{sec:1-1}
The fluid is  an electrolyte solution of the solvent and two ionic species with different {valencies $\zi$, $\valaeq, z_1=-1, z_2=+1$.}
The ionic transport in the fluid is driven by three phenomena: 1) convection of the solvent which is determined by a convective velocity $\wb$,  2) diffusion of the $\vala$-th ionic species in the solvent characterized by the diffusivity $D_\vala$ of the $\vala$-th ionic species, and 3) motion of ions due to the electrical field. 
In what follows, we introduce the system of equations describing these processes in the pore space $\Omega_f$ filled by the electrolyte solution. All the electrochemical and mechanical constants involved in the model are defined in Tab.~\ref{tab_constants}.

Each species (labeled by $\alpha$) dissolved in the electrolyte is associated with the electrochemical potential $\mui$, which depends on the concentration $\ci$, the electrostatic potential $\Psi$, and temperature $T$, such that , see \cite{hunter2001foundations},
\begin{equation}\label{eq_elchem_potential}
\mui=\mui^0+k_B T\ln \ci+e\zi\Psi\;,
\end{equation}
where $\mui^0$ is the standard electrochemical potential expressed at infinite dilution, $k_B$ is Boltzmann constant, and  $e$ is the elementary charge. The transport is restricted to the Eulerian mass conservation law, 
\begin{equation}\label{eq_balance}
\frac{\pd \ci}{\pd t} + \diver \left(\jbi+\wb \ci\right)=0\qquad\textrm{ in }\Omega_f, \valaeq,
\end{equation}
where $\wb$ stands for convective velocity and the effects of diffusion and migration caused by an external electrical field are expressed by the migration-diffusion flux $\jbi$ given by 
\begin{equation}\label{eq_j}
\jbi=-\frac{\ci D_\vala}{k_BT}\nabla\mu_\vala=-\frac{\ci D_\vala}{k_BT}(\frac{k_BT}{\ci }\nabla \ci  + e\zi\nabla\Psi),\qquad \valaeq,
\end{equation}
whereby, the no-flux condition holds on the solid-fluid interface $\Gamma$,
\begin{equation}\label{eq_j_bc}
\jbi \cdot\nb=0\qquad\textrm{ on }\Gamma, \valaeq.
\end{equation}

The electrokinetics of the fluid phase is characterized by distribution of the electrostatic potential $\Psi$ which satisfies the Gauss-Poisson problem,
\begin{equation}\label{eq_elektrokinetic}
\mathcal{E}\nabla\cdot\Eb =e\sum\limits_{\valblim}^{2}\zj\cj\inomf,\quad \Eb = -\nabla\Psi\;,
\end{equation}
where $\mathcal{E}$ is dielectric coefficient of the solvent (assumed to be constant), and $\mathbf{E}$ is the  electric field. 

Obviously, the fluid flow is governed by the Navier-Stokes equations involving the fluid velocity $\vb$, the hydrostatic pressure $p$. Since we consider slow flows only, the convective term can be neglected as well as the inertia term, thus, the fluid stress tensor $\boldsymbol{\boldsymbol{\sigma}}_f$ satisfies the equilibrium equation,
\begin{equation}\label{eq_fluid}
-\diver\boldsymbol{\sigma}_f=\fb\inomf,
\end{equation}
where $\fb$ is the external body force. The stress in the fluid phase is extended by the Maxwell 2nd order stress tensor $\boldsymbol{\tau}_M=\mathcal{E}\left(\Eb\otimes\Eb-\frac{1}{2}|\Eb|^2\Ib\right)$ so that
\begin{equation}\label{eq_stress_fluid}
\boldsymbol{\sigma}_f=-p\Ib+2\eta_f\strv+\boldsymbol{\tau}_M,
\end{equation} 
where  $\strv=\frac{1}{2}\left(\nabla\vb+(\nabla\vb)^{\textrm{T}}\right)$, $p$ is the fluid pressure and $\eta_f$ is the dynamic viscosity of the electrolyte. The flow is assumed to be incompressible, hence
\begin{equation}\label{eq_incompressibility}
\diver\vb=0\inomf.
\end{equation}
Upon substituting \Eq{eq_stress_fluid} into \Eq{eq_fluid}, the Poisson equation \Eq{eq_elektrokinetic} allows to rewrite the term $\diver{\boldsymbol{\tau}_M}$ in terms of $\nabla\Psi$, such that the modified Stokes problem is obtained,
\begin{equation}\label{eq_stokes}                           
\nabla p-\eta_f\nabla^2\vb=\fb-e\sum\limits_{\valblim}^{2}\zj\cj\nabla\Psi\inomf.
\end{equation}
Let us note that if the porous medium is considered to be rigid, the influence of solid matrix on fluid phase can be expressed only by boundary conditions on  $\Gamma$. 
However, in case of the deformable porous body, the model needs to be expanded by equations describing processes in solid phase. 

\subsection{Fluid solid interaction}
\label{sec:1-2}
\paragraph{Model of the solid phase} We consider solid skeleton of the porous medium is constituted by an elastic conducting material. Therefore, as stated above, we assume the electric potential $\Psi_s$ is constant in the whole skeleton $\Om_s$, whereas a constant surface charge $\Sigma$ is distributed on its surface $\pd \Om_s$. Under standard assumptions of linear elasticity, small displacement and deformation, the displacement field $\ub$ satisfies 
\begin{equation}\label{eq_elasticity}
-\diver\boldsymbol{\sigma}_s=\fb\inoms,\quad \boldsymbol{\sigma}_s=\Ab\stru,
\end{equation}
 where $\stru=\frac{1}{2}\left(\nabla\ub+(\nabla\ub)^{\textrm{T}}\right)$ is  the strain tensor,  and the elasticity tensor $\Ab=(A_{ijkl})$ is symmetric and positive definite, $A_{ijkl}=A_{jikl}=A_{klij}$.

 \paragraph{Interface conditions}
 To complete the model of the fluid saturated porous medium, the interface conditions must be prescribed on $\Gamma$.
It should be pointed out that the transmission condition on  $\Gamma$ concerning the electric field or electric potential may be defined using two methods.
While considering no-slip hydrodynamic condition on the solid surface, \ie $\vb= \dot\ub$, the electric potential on  $\Gamma$ can be introduced using the $\zeta$-potential. This parameter is widely used in definition of the EDL, but it is more related to the electrochemical properties of the system, \cite{hunter2001foundations}.
{However, to treat more general situations}, a surface charge density $\Sigma$ proportional to the normal derivative of $\Psi$ is often used instead of prescribing the $\zeta$-potential. Since, in our setting, potential $\Psi_s$ is a constant,  the surface charge $\Sigma$ given at  $\Gamma$ constitutes the boundary condition,
\begin{equation}\label{eq_elektrokinetic_bc}
\mathcal{E}\Eb\cdot\nb=\Sigma\ongamma.
\end{equation}

Concerning the mechanical interaction on  $\Gamma$, in general, the convective velocity $\wb$ takes into an account the solid deformation extending to the fluid part and $\wb=\vb-\dot\ub$.
In this paper, we restrict to stationary problems so that the fluid velocity is equivalent to the convective velocity,
\begin{equation}\label{eq_w}
\wb=\vb.
\end{equation}
Moreover, by virtue of the linear kinematics (small displacements and deformations) only ``one-way'' interaction can be considered: the solid is loaded by traction forces due to the fluid stress, however the walls of the pores $\Om_f$ are considered as rigid for the flow model. Therefore, the following   interface conditions  ensure the no-slip of the flow and the continuity of the normal stresses,
\begin{eqnarray}
\wb=0\ongamma,\\
\boldsymbol{\sigma}_f\cdot\nb=\boldsymbol{\sigma}_s\cdot\nb\ongamma.\label{eq_elasticity_bc}
\end{eqnarray}

%

\begin{table}[h]\centering
\begin{tabular}{llll}
\hline\noalign{\smallskip}
		        	Symbol	& Quantity &Value&Unit\\
\noalign{\smallskip}\hline\noalign{\smallskip}
		        $e$&Electron charge&$1.6\times 10^{-19}$&	C	       \\
		        $k_B$&Boltzmann constant&$1.38\times 10^{-23}$&	J/K\\  
		        $T$&Absolute temperature&$298$&	K\\	       
		        $\mathcal{E}$&Dielectric constant&$6.93\times 10^{-10}$& c/(mV)	\\	        
		        $\eta_f$&Dynamic viscosity of fluid&$1\times 10^{-3}$&kg/(ms)\\
		         $D_1$&Diffusivity of 1st ionic species&$13.33\times 10^{-10}$&m$^2$/s\\
		          $D_2$&Diffusivity of 2nd ionic species&$20.32\times 10^{-10}$&m$^2$/s\\
		           $l$&Characteristic pore size&$1.0\times 10^{-7}$&m\\
		           $c_c$&Characteristic concentration&$(6.02\times 10^{24},6.02\times 10^{26})$& particles/m$^3$\\
		           $\Sigma_c$&Surface charge density&$-0.129$&C/m$^2$\\
		           $\Lambda$&Young modulus&$7.3\times 10^{9}$&Pa\\
\noalign{\smallskip}\hline
		        \end{tabular}
		        \caption{Description of used parameters, source \cite{allaire2015ion}.}
		        \label{tab_constants}
\end{table}

\paragraph{Model restrictions}
In this paper, we restrict the modelling to stationary problems, so that the time derivative $\pd_t\ci$ in \Eq{eq_balance} vanishes. As the result, the fluid-structure mechanical interaction simplifies and also the electrochemical interactions in the fluid can be treated using the assumption of the electroneutrality in the bulk electrolyte.
{Finally,  to define boundary conditions on the exterior boundary $\partial\Om$. we shall assume that $(\Psi+\Psiex), c_i, \ub, \wb $ and $p$ are $L-$periodic, where $L$ is the  side of the cube $\Om = ]0,L[^d$.}


\section{Linearization and decomposition into subproblems}\label{sec:lin}
In order to apply the homogenization method, the complex nonlinear problem involving electrochemical and fluid-structure interactions is linearized. We introduce the equilibrium state which enables to establish a linearized problem for perturbation fields. The electrostatic problem is related to the heterogeneous periodic structure at the pore level.

\subsection{Periodic structure of the porous medium}
The porous medium is characterized by parameter $\ell^\mic$  characterizing the size of the micropores. By virtue of the  upscaling, the characteristic length $\ell^\mic$ is related to  a given macroscopic characteristic length $L$, such that the scale parameter introduced by $\veps = \ell^\mic/L$.

The porous medium  is generated as a periodic lattice by repeating the representative volume element (RVE) occupying domain $Y^\veps = \veps Y$, see \ref{sec_A}.
According to the decomposition introduced in Section~\ref{sec:1}, the zoomed cell 
$Y = \Pi_{i=1}^3]0,\bar y_i[ \subset \RR^3$ 
splits into the solid part
occupying domain $Y_s$ and the complementary fluid part $Y_f$, thus
\begin{equation}\label{eq-mi6}
\begin{split}
Y   = Y_s  \cup Y_f  \cup \Gamma_Y \;,\quad
Y_s   = Y \setminus \ol{Y_f}  \;,\quad
\Gamma_Y   = \ol{Y_s } \cap \ol{Y_f }\;.
\end{split}
\end{equation}
Note that, roughly speaking, the relation between the micro- and macroscopic coordinates is $y=x/\veps$, see the precise decomposition ansatz in \Eq{eq:3a}.
For a given scale $\veps > 0$, $\ell_i = \veps \bar y_i$ is the characteristic size associated with the $i$-th coordinate direction, whereby
also $\veps \approx \ell_i / L$, hence $\ell_i \approx \ell^\mic$ (for all $i=1,2,3$) specifies the microscopic characteristic length $\ell^\mic$. 
Below we introduce two-scale functions depending on $x \in \Om$ and $y \in Y$ using the unfolding operator


 \begin{figure}[t]
 	\centering
 	\includegraphics[width=10cm]{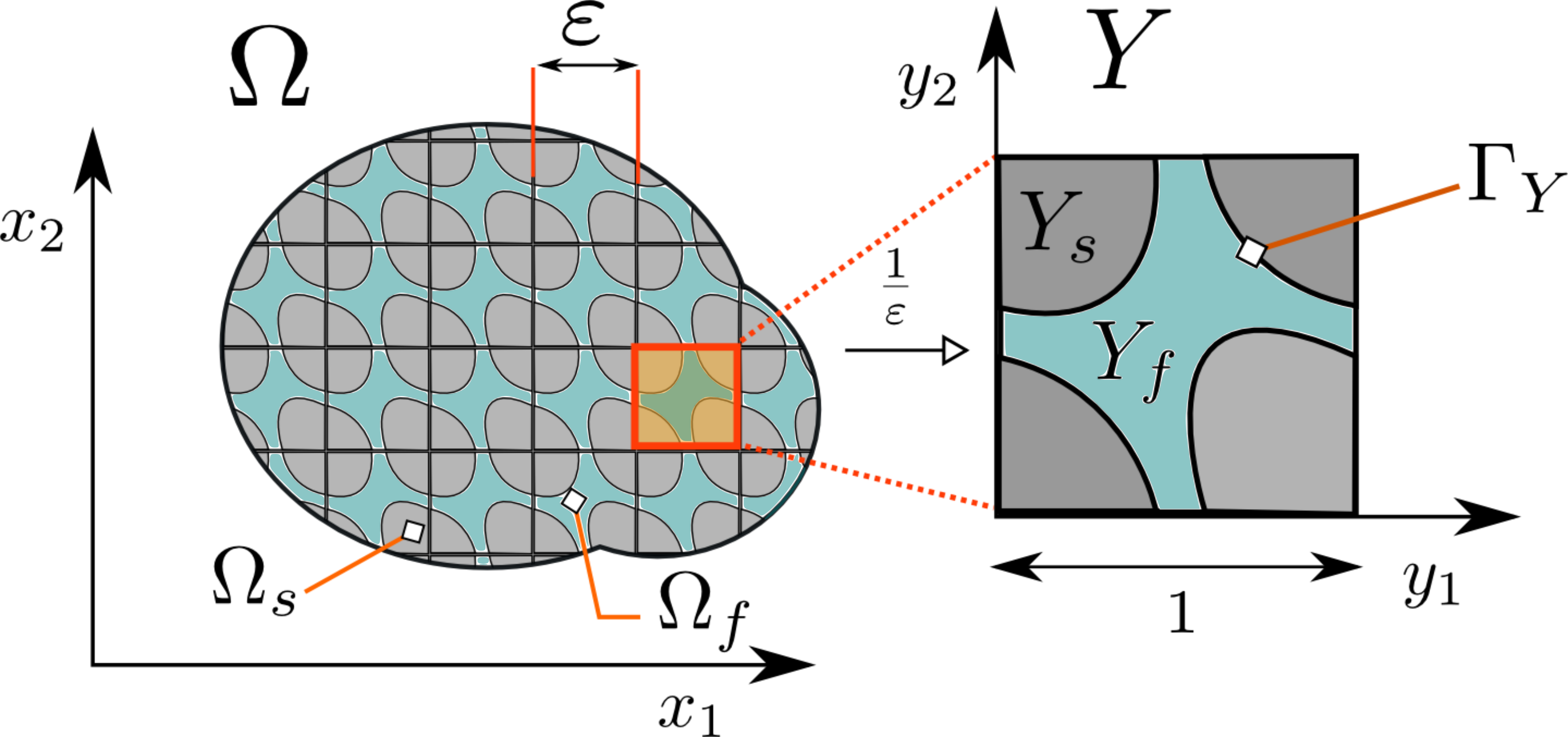}
 	\caption{Domain $\Om$ is generated by periodical repeating of representative periodic cell $Y$.}
 	\label{fig_domain}
 \end{figure}
 
 The ratio between the characteristic dimension of macroscopic domain $L_{c}$ (this may be associated with the specimen size $L$) and characteristic dimension of microscopic periodic structure $l$ is a scale parameter  $\varepsilon=\frac{l}{L_{c}}, 0<\varepsilon\ll 1$  and represents the smallest zoom, by which the microstructure becomes visible from the macroscopic point of view.

\subsection{Dimensionless problem}\label{sec:dimless}
In this section we shall introduce a non-dimensional form of the equations from Section~\ref{sec:1-1} and \ref{sec:1-2}, following the approach from \cite{allaire2015ion}. The macroscopic coordinate is rescaled by $x^\prime=x/L_c$ using the characteristic length $L_c$.
Accordingly, the dimensionless operator $\nabla^\prime$ is defined by
\begin{equation}\label{dimless_o}
\nabla^\prime=(\partial_{x^\prime})=L_c(\partial_x)=L_c\nabla.
\end{equation}
In the following text, however, we shall drop the prime $^\prime$, to simplify the notation, so that by $x$ we refer to the rescaled macroscopic coordinates.
The dimensionless variables are expressed in terms of the characteristic quantities denoted by subscript $\sqcup_c$. As they are related to scale parameter $\veps$, we denote them all by superscript $\sqcup^\veps$.
\begin{equation}\label{dimless_q}
\pe=\frac{p}{p_c}\;,\quad \wbe=\frac{\wb}{v_c}\;,\quad {\Psie=\frac{\Psi}{\Psi_c}}\;, 
\quad\cie=\frac{\ci }{c_c}\;,\quad\ube=\frac{\ub}{u_c},
\end{equation}
where $p_c$ is the characteristic pressure, $v_c$ the characteristic velocity, $c_c$ characteristic concentration, $\Psi_c = k_B T/ e$ the characteristic potential, $u_c$ the characteristic displacement. 
Further, the dimensionless forcing terms labeled  by $\sqcup^*$ are defined,
 \begin{equation}\label{dimless_f}
 \Psieh=\frac{\Psi^{\textrm{ext}}(x)}{\Psi_c}\;,\quad\Eb^*=\nabla\Psieh \;,\quad\Sigmah=\frac{\Sigma}{\Sigma_c}\;,\quad\fbh=\frac{\fb L_c}{p_c}\;.
 \end{equation}
 
The thickness of the electric double layer (EDL) is represented by the  Debye length $\lambda_D$,
\begin{equation}\label{debye}
\lambda_D=\sqrt{\mathcal{E}\frac{\Psi_c}{e c_c}\left(\sum\limits_{\valblim}^{2}\zj^2\right)^{-1}}
=\sqrt{\mathcal{E}k_BT\left(e^2c_c\sum\limits_{\valblim}^{2}\zj^2\right)^{-1}} = ,
\end{equation}
where the last expression is due to the special type of the electrolyte with $z_1=-1, z_2=+1$.
 Apart of this parameter,  other two dimensionless parameters are employed in the  the dimensionless form of the system \Eq{eq_balance}-\Eq{eq_incompressibility}; these are the Peclect number $ \Peci=\frac{l^2k_BTc_c}{\eta_f D_\vala}$, and the ratio between electrical and thermal energy $\Nsig=\frac{el\Sigma_c}{\mathcal{E}k_BT}$.   
 
 \paragraph{Nondimesionalized problem} Using
   the dimensionless form of the system \Eq{eq_balance}-\Eq{eq_incompressibility} we can state the following problem. Given $\fb^*,\Eb^*, \Sigma^*$, find $c_\alpha^\veps, \Psi^\veps, p^\veps, \wb^\veps$ and $\ub^\veps$ which satisfy the following set of equations,
	\begin{align}\label{eq_adim_start}
	\veps^2\nabla^2\wbe-\nabla\pe&=\fb^*+\sum_{\valblim}^{2}\zj\cjex\nabla\Psie&\inome,\\
	\diver\wbe&= 0&\inome,\\
	\veps^2\nabla^2\Psie&=\betapar \sum_{\valblim}^{2}\zj\cjex&\inome,\label{eq_adim_PB1}\\
	\nabla\cdot\left( \Peci\wbe_\vala\ci^\veps+\jbie\right)&= 0&\inome, \valaeq,\\
	-\nabla\cdot(\Ab^*\strue)&=\fbh&\quad\textrm{ in }\Omega^\veps_s, \label{eq_deform_adim_start} 
	\end{align}
	with interface conditions 
	\begin{align}
	\wbe&=0&\onome,\\
	\jbie\cdot\nb&= 0&\onome,\valaeq,\\
	\veps\nabla\Psie\cdot\nb&=-\Nsig\Sigma^*&\onome,\label{eq_adim_stop}\\
	\Ab^*\strue\cdot\nb&=\sigmabf_f^\veps\cdot\nb&\onome,\label{eq_deform_adim_stop}
	\end{align}
		where $\betapar=\frac{l^2}{\lambda_D^2\sum_{\valblim}^{2}\zj^2}$ and
		\begin{align}
		\jbie =&-\cie\nabla\left(\ln\cie\exp(z_i\Psie)\right),\label{eq-j-grad}\\
		\sigmabf_f^\veps=&- p^\veps\Ib+2\veps^2\str{\wbe}+\betaparinv\veps^2\left(\nabla\Psie\otimes\nabla\Psie-\frac{1}{2}|\nabla\Psie|^2\Ib\right).
		\end{align}
 Concentrations $\cje$, sum of potentials $\left(\Psie+\Psieh\right)$, convective velocity $\wbe$  and pressure $\pe$ are {$L$-periodic.  }
We also recall that the given surface charge density $\Sigma^*$ is constant, as the consequence of the assumed constant potential $\Psi_s$ in the solid conductor. Introduction of the parameter $\veps$ into \Eq{eq_adim_start},\Eq{eq_adim_PB1} and \Eq{eq_adim_stop} is the natural consequence of the adimensional choices and dimensional analysis of the system, see \ref{sec_B}.

 \subsection{Linearization}
 To apply the homogenization efficiently, the system \Eq{eq_adim_start}-\Eq{eq_deform_adim_stop} must be linearized. For this, the assumption of sufficiently small applied fields  $\Psieh$ and $\fb^*$ is needed. As the consequence, the state variables are only slightly perturbed from equilibrium, which justifies the linearization.

 Following the linearization procedure in \cite{allaire2013ion}, any unknown $a^\veps$ can be
decomposed into its equilibrium part $a^\eqe$ and its perturbation $\delta a^\veps$, thus, we consider 
	\begin{equation}
	\begin{split}\label{eq_decomposition1}
	\ciex=\cioex+\dciex,& \qquad\Psie(x)=\Psioe(x)+\dPsie(x),\\
	\wbe(x)=\wboe(x)+\dwbe(x),& \qquad\pe(x)=\poe(x)+\dpe(x),\\
	\ube(x)=\uboe(x)+\dube(x),&
	\end{split}
	\end{equation}
  The equilibrium quantities labeled superscript $\sqcup^\textrm{eq}$, are solutions of the system \Eq{eq_adim_start}-\Eq{eq_deform_adim_stop} for $\fb{^*}=0$, $\Psieh=0$, $\wboe=0$  and, by the consequence,  zero diffusive fluxes $\jbie = 0$; the last statement follows from \Eq{eq_balance}, \Eq{eq_j_bc}, and \Eq{eq-j-grad}.
 Since the convective velocity vanishes at the equilibrium, we can state $\wbe(x)=\dwbe(x)$. 

 \subsubsection{Equilibrium state quantities}
 Obviously, the equilibrium solution defines the reference state of the electrolyte such that the linearized state problem governs the perturbations. 
 The existence of equilibrium solution $(\cioex,\Psioe,\poe,\uboe)$ was shown in \cite{allaire2013asymptotic}, whereby the three fields $\cioex,\Psioe$ and $\poe$ satisfy the following two relationships,
 \begin{equation}\label{eq_equilibrium_conc}
   \begin{split}
   \qquad\cioex & = \cib\exp(-\zj\Psioe(x)),\\
   \poe(x) & =\sum_{\valblim}^{2}\cjoex ,
    \end{split}
 \end{equation}
 where by $\cib$ is the characteristic concentration in the bulk which represents the concentration of the $\vala$-th ionic species is an infinite pore. As the consequence of \Eq{eq_equilibrium_conc}, $\Psioe$ determines $\cioex$, then $\poe$.
 
 To compute $\Psioe$,  the asymptotic analysis of the dimensionless Poisson-Boltzmann equation given by \Eq{eq_adim_PB1} and \Eq{eq_adim_stop} has been treated in \cite{allaire2013asymptotic}. Therefore, for the sake of completeness, here   we only provide the resulting expressions. 
 Substituting  equilibrium concentration \Eq{eq_equilibrium_conc}$_1$ into the Poisson-Boltzmann equation \Eq{eq_adim_PB1} and \Eq{eq_adim_stop}, one gets
 \begin{equation}
  \begin{split}
 \veps^2\nabla^2\Psioe=&\betapar\sum_{\valblim}^{2}\zj\cjb\exp(-\zj\Psioe)\quad\inome,\label{eq_adim_PB}\\
 \veps\nabla\Psioe\cdot\nb=&-\Nsig\Sigma^*\quad \onome,
 \end{split}
  \end{equation}
 where the $L$-periodicity of $\Psioe$ is prescribed on the external boundary $\pd_\ext\Om_f^\veps$. The solvability of \Eq{eq_adim_PB} for the zero Neumann condition, $\Sigma^*=0$, requires that the \rhs integrated in $\Om_f^\veps$ must vanish. This is satisfied provided the so-called electroneutrality condition in bulk holds,
 \begin{equation}\label{eq-elneutr}
 \sum_{\valblim}^{2}\zj\cjb=0.
 \end{equation}
 We adhere this condition, hence the existence of a unique solution $\Psioe\in H^1_\#(\Om^\veps_f)$ is guaranteed. 
 From the physical point of view it ensures that $\Psioe$  vanishes for the zero surface charge.

 Although we assume $\Sigma^*$ to be a constant defined on interface $\Gamma^\veps$, even for a periodic distribution of charges $\Tuf{\Sigma^*}=\tilde\Sigma^*(y)$, $ y\in \Gamma_Y$, the problem \Eq{eq_adim_PB} yields $\veps Y$-periodic solutions $\Psioe$ in $\Om_f^\veps$, recalling the ``macroscopic'' $L$-periodicity on $\pd_\ext\Om_f^\veps$. This property allows us to consider only the local problem in the zoomed RVE represented by cell $Y_f$. 
 Then,
 \begin{equation}
 	\Psioe(x)=\Psieq(y), \quad  \cioe(x)=\cio(y),
 \end{equation}
 where concentrations $\cio(y), \vala=1,2$ obey the form of the Boltzmann distribution
 \begin{equation}\label{eq_concentration}
 \cio(y)=\cib\exp(-\zi\Psieq(y)).
 \end{equation} 
 Potential $\Psieq(y)\in H^1_\#(Y_f)$ is a solution of the Poisson-Boltzmann equation \Eq{eq_adim_PB} imposed in  $Y_f$, in particular 
 \begin{equation}
  \begin{split}
 \nablay^2\Psieq=&\betapar\sum_{\valblim}^{2}\zj\cjb\exp(-\zj\Psieq)\inyf,\label{eq_adim_PB0}\\
 \nablay\Psieq\cdot\nb=&-\Nsig\Sigma^*\ongammay.
 \end{split}
  \end{equation}
  
 \begin{figure}[!h]
 	     \centering
 		\includegraphics[width=10cm]{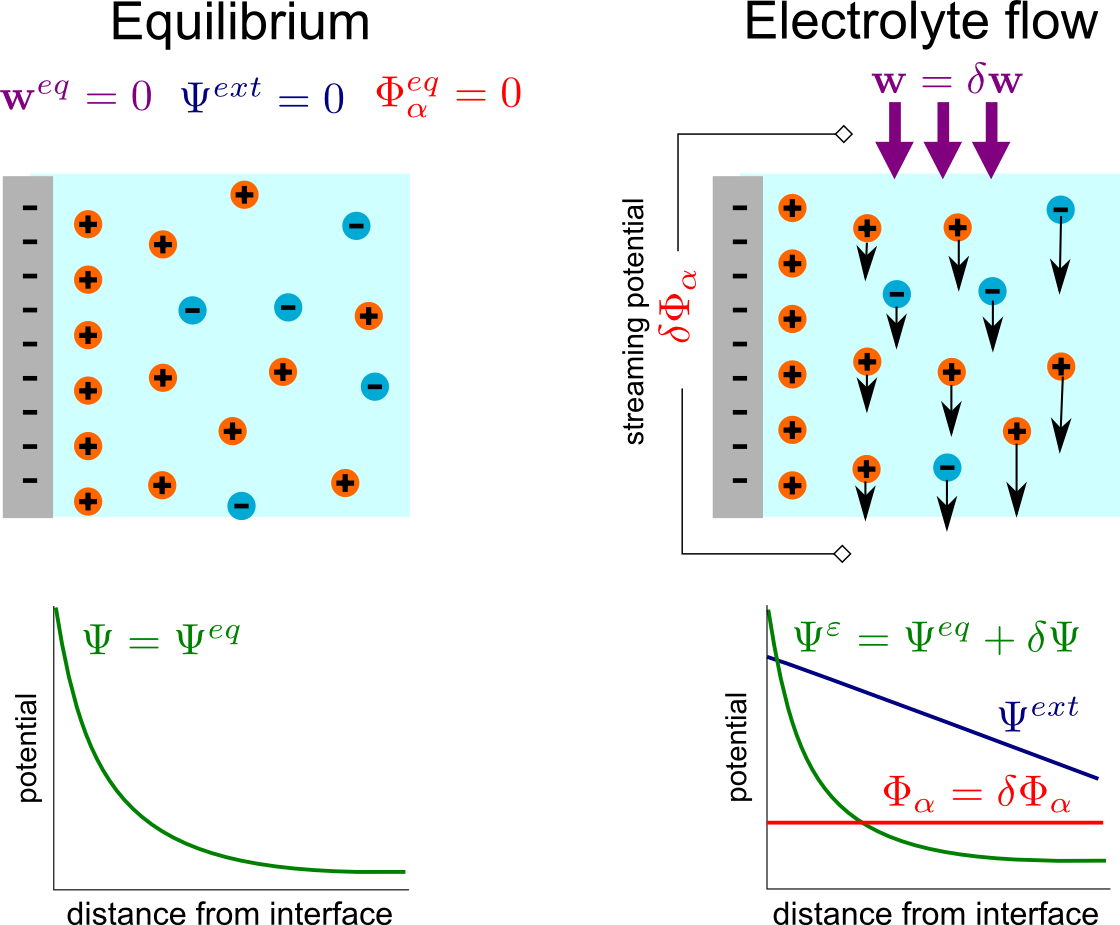}
 	     \caption{Potential decomposition near solid-fluid interface in the equilibrium and non equilibrium state. Occurrence of streaming potential under the electrolyte flow.}
 	     \label{fig_edl}
 \end{figure}

 To conclude, by virtue of \Eq{eq_equilibrium_conc}, the unfolded  equilibrium concentrations $\Tuf{\cioe}$ and the unfolded pressure field $\Tuf{\poe}$ are $Y$-periodic functions. Moreover, by the consequence, the unfolded displacements are also $Y$-periodic functions, whereby the macroscopic strains vanish. Therefore, in this paper, we neglect any influence of the  equilibrium displacements field on the reference configuration associated with the linearization procedure considered in what follows. Note that, the equilibrium pore geometry might be perturbed due to the local strains in $Y_s$. 

 \subsubsection{Perturbed state quantities}
 As the further step in the linearization, the total electrostatic potential $\Psi_f^\veps$ is decomposed according to phenomena which participate in the total electric field. This  can be considered as a superposition $\Psi_f^\veps=\Psie+\Phie+\Psiex$ of local particular electric fields associated with potentials $\Psie$, $\Phie$, and $\Psiex$
\begin{itemize}
\item potential  $\Psiex$ which yields the external electrical field $\Eb^* = -\nabla \Psiex$ is imposed and independent of $\veps$; 
\item potential $\Psie$ reflects only the effects of the EDL on the ion distribution; in the equilibrium state, $\Psie$ is given by problem \Eq{eq_adim_PB};
\item ionic potentials $\Phi_\vala$ (often referred to as the streaming potential) represents the electric field produced by  motion of $\vala-$th ionic species. In the equilibrium, $\Phie$ vanishes since both the convection $\wbe$ and the ionic flux $\jbie$ vanish. Thus, the ionic potential is identified by its perturbation only, $\Phie(x)=\dPhie$.

\end{itemize}
To summarize the decomposition, the potential in equilibrium is given only by the electrokinetic potential, so that $\Psi^{\rm{tot,eq}}=\Psieq$. The decomposition of the potential in equilibrium and under the flow is also illustrated in Fig.\ref{fig_edl}.

  Ionic concentrations can be expressed in the context of Boltzmann distribution, so that
  \begin{equation}\label{eq_ionic_pot}
  \cie(x)=\cib\exp(-\zi\left(\Psie(x)+\Phie(x)+\Psieh(x)\right)).
  \end{equation}
  The introduction of ionic potentials will help to eliminate the boundary condition \Eq{eq_adim_stop} from the system, because the mobility of particles is not influenced by the choice of boundary conditions, see \cite{obrien1978electrophoretic}. The boundary condition \Eq{eq_adim_PB}$_2$ is considered only to define $\Psi^\eqe$, the potential distribution in equilibrium. It will be shown later that potential $\Phie$ proves useful in the decoupling the electrokinetic system.

  The linearization of \Eq{eq_ionic_pot} by the first-order Taylor expansion yields
 \begin{equation}\label{eq_delta_c}
 \dcie(x)=-\cioe(x)\zi\left(\dPsie(x)+\dPhie(x)+\Psieh(x)\right).
 \end{equation}
 Further, following the work \cite{moyne2006two}, it is convenient to introduce the so-called global pressure $P^\veps$,
  \begin{equation}\label{eq_P}
P^\veps=\dpe+\sum\limits_{j}^{2}\cjoe \zj \left(\dPsie+\dPhije+\Psieh\right)\;,
 \end{equation}
  which consists of the hydrodynamic pressure perturbation $\dpe$ and the osmotic pressure, see \cite{lemaire2010multiphysical}.

  Finally, the decomposition of unknowns fields \Eq{eq_decomposition1} is substituted into the dimensionless problem \Eq{eq_adim_start}-\Eq{eq_deform_adim_stop} and  \Eq{eq_delta_c} is employed to express $\dcie(x)$.
  Note, that products of the small quantities
 such as $\dcie\dPsie$  or $(\dcie)^2$ are neglected.
  Due to the linearization and the use of the global pressure \Eq{eq_P}, the nondimensionalized problem splits into three subproblems which can be solved subsequently. 
  
  \paragraph{Linearized electrokinetic system}
  Given the body forces $\fb^*$ and potential $\Psieh$, find $L$-periodic functions $(\dwbe, P^\veps, \dPhie, \dPsie)$ and $\dube$ which solve the following three subproblems:
  \begin{enumerate}

  \item  Electrokinetic problem:   $(\dwbe, P^\veps, \dPhie)$ satisfy
 \begin{align}\label{eq_lin_start1}
 \veps^2\nabla^2\dwbe-\nabla\Pe=-\fb^*-\sum_{\valblim}^{2}\zj\cjoex&(\nabla\dPhije+\nabla\Psieh)&\inome,\\
 \diver\dwbe&=0&\inome,\\
 \diver\left(\cioex\left(\nabla\dPhie+\nabla\Psieh+\frac{\Peci}{\zi}\dwbe\right)\right)&=0&\inome, \label{eq_lin_balance} \\
  \dwbe&=0&\onome,\label{eq_lin_w_bc}\\
 (\nabla\dPhie+\nabla\Psieh)\cdot\nb&=0&\onome.\label{eq_lin_bc_phi}
\end{align}

\item Electrostatic EDL problem: $\dPsie$ satisfies
  \begin{align} 
 -\veps^2\nabla^2\dPsie + \betapar\left(\sum_{\valblim}^{2}\zj^2\cjoex\right)\dPsie=-\betapar&\sum_{\valblim}^{N}\zj^2\cjoex(\Phi^\veps_\valb+\Psieh)&\inome,\label{eq_lin_pot1}\\ 
\nabla\dPsie\cdot\nb&=0&\onome.\label{eq_lin_pot2}
 \end{align}

\item Deformation problem: $\dube$ satisfies
\begin{align}
   -\diver(\Ab^*\strdue)&=\fb^*&\textrm{ in }\Omega^\veps_s,\label{eq_lin_disp1} \\
 (\Ab^*\strdue)\cdot\nb&=\sigmabf_f^\veps\cdot\nb &\onome,\label{eq_lin_stop}
 \end{align}
 with the linearized fluid stress given by
 \begin{align}
 \sigmabf_f^\veps=-\Pe\Ib+2\veps^2\str{\dwbe}+\sum\limits_{\valblim}^{2}\zj\cjoe(\dPsie+\dPhije+\Psieh)\Ib\nonumber\\
 +\betaparinv\veps^2\left(\nabla\Psioe\otimes\nabla\dPsie+\nabla\dPsie\otimes\nabla\Psioe-\nabla\Psioe\cdot\nabla\dPsie\Ib\right)\label{eq_fluid_stress_lin}.
 \end{align}
 \end{enumerate}
 It is worth to note that, due to the linearization and decoupling in three subproblems,  \Eq{eq_lin_w_bc}-\Eq{eq_lin_bc_phi}, \Eq{eq_lin_pot2} and \Eq{eq_lin_stop} present stanard boundary conditions on interface $\Gamma^\veps$, rather than transmission conditions, as in problem \Eq{eq_adim_start}-\Eq{eq_deform_adim_stop}.

 \section{Homogenization}\label{sec_homog}
 The unfolding homogenization method \cite{cioranescu2008periodic} has been used for the asymptotic analysis $\veps\rightarrow 0$ of  weak formulation \Eq{eq_weak_elkin1} arising from the linear system \Eq{eq_lin_start1}-\Eq{eq_lin_stop}.
 Since the resulting system of the limit two-scale equations  corresponds to the one obtained in paper \cite{allaire2013ion}, in the next sections, we report only briefly the on the upscaling procedure. The main purpose is to explain the structure of the homogenized model, namely to define local problems for the so-called characteristic responses, to give formulae for computing the homogenized coefficients, and to formulate the macroscopic problem.  
 
 \subsection{Convergence results}\label{sec_conv}
 The convergence analysis is derived for the weak formulation of the linearized problem \Eq{eq_lin_start1}-\Eq{eq_lin_stop}.
 The following functional spaces will be employed.
 \begin{align}
 \Hb^1_{\#}(\Om_p^\veps)=&\left\{\phibf\in H^1(\Om_p^\veps)^d,  \ L-\textrm{periodic in }x \right\},\nonumber\\
 \Hb^1_{\#0}(\Om_p^\veps)=&\left\{\phibf\in H^1(\Om_p^\veps)^d,  \phibf=0\ongamma^\veps,\ L-\textrm{periodic in }x \right\},\nonumber\\
 H^1_\#(\Om_p^\veps)=&\left\{\psi\in H^1(\Om_p^\veps),\ L-\textrm{periodic in }x\right\},\nonumber
 \end{align}
 where $H^1(\Om_p^\veps)$ is the Sobolev space $W^{1,2}(\Om_p^\veps)$, subscript $p=s,f$.

 \paragraph{Weak formulation of the linearized electrokinetic system}
 Given $\fb^* \in \Lb^2(\Om^\veps_f)$ and $\Eb^* \in \RR^3$, whereby $\Psieh =  x\cdot \Eb^*$.
  \begin{enumerate}

  \item Find $(\dwbe,\Pe,\dPhie)\in \Hb^1_{\#0}(\Om^\veps_f)\times [H^1_\#(\Om^\veps_f)]^3$, such that
 \begin{align}\label{eq_weak_elkin1}
 \veps^2\int\limits_{\Omega_f^\veps}\nabla\dwbe:\nabla\testw \dx+\int\limits_{\Omega_f^\veps}\testw\nabla\Pe \dx-\sum_{\valblim}^{2}\zj&\int\limits_{\Omega_f^\veps}\cjoe\testw\cdot(\nabla\dPhije+\Eb^*)\dx=\nonumber\\
 &=\int\limits_{\Omega_f^\veps}\testw\cdot\fb{^*}\dx, \qquad\forall\testw\in\Hb^1_{\#0}(\Omega_f^\veps),\\
 \int\limits_{\Omega_f^\veps}\cioe\nabla\dPhie\nabla\testphi \dx +\int\limits_{\Omega_f^\veps}\cioe\frac{\Peci}{\zi}\dwbe\nabla\testphi \dx&=\nonumber\\
  =-\int\limits_{\Omega_f^\veps}\cioe\Eb^*\cdot\nabla\testphi \dx,& \qquad\forall\testphi\in H^1_\#(\Omega_f^\veps), \valaeq,\\
    \int\limits_{\Omega_f^\veps}\testp \nabla\cdot\dwbe\dx&=0,\qquad\forall\testp\in H^1_\#(\Omega_f^\veps)\label{eq_weak_elkin2}
 \end{align}
 
 \item Find $\dPsie\in H^1_\#(\Om^\veps_f)$, such that
 \begin{align}
 \veps^2\int\limits_{\Omega^\veps_f}\nabla\dPsie\cdot\testpsi \dx +& \betapar \int\limits_{\Omega^\veps_f}\left(\sum_{\valblim}^{2}\zj^2\cjoe\right)\dPsie\testpsi \dx = \nonumber\\
 =&-\betapar\sum_{\valblim}^{2}\zj^2\int\limits_{\Omega^\veps_f}\cjoe\left(\dPhije+\Psieh\right)\testpsi \dx,\qquad\forall\varphi\in H^1_\#(\Om^\veps_f).
 \end{align}
 \item Find $\ube\in H^1_{\#}(\Om_s)^d$, such that
 \begin{equation}\label{eq_weak_disp}
 \int\limits_{\Omega^\veps_s}\Ab^*\strdue\eb(\testu)\dx =\int\limits_{\Gamma^\veps}\sigmabf^\veps_f\nb\cdot\testu\dS+\int\limits_{\Omega^\veps_s}\fb^*\cdot\testu\dx, \qquad\forall\testu \in H^1_{\#}(\Om^\veps_s)^d,
 \end{equation}
 \end{enumerate}
 where $\sigmabf^\veps_f$ is given by \Eq{eq_fluid_stress_lin}.

  The two-scale limit problem can be obtained due to the weak convergences in the unfolded domain $\Om\times Y$; the two-scale convergence method was used in  \cite{allaire2010homogenizationA} and \cite{allaire2015ion}. The unfolded equations of the weak formulation are obtained using the unfolding operator $\mathcal{T}_\veps$ defined in  \ref{sec_A}, see \cite{cioranescu2008periodic}.
  Due to the a~priori estimates on the solutions of \Eq{eq_weak_elkin1}-\Eq{eq_weak_disp}, according to 
 \cite{allaire2015ion} the following convergence result for $\veps\rightarrow 0$ can be proved:
 There exist limit fields $(\dwb^0, P^0)\in L^2(\Omega;H^1_\#(Y_f)^d)\times L^2(\Omega)$, $\left\{\dPhi^0_\valb,\dPhi^1_\valb\right\}_{\valblim,\dots,N}\in (H^1(\Omega)\times L^2(\Omega;H^1_\#(Y_f)))^N$, $\dPsi^0\in L^2(\Omega;H^1_\#(Y_f))$, $\dub^0\in H^1_{\#}(\Om_s^\veps)^d$ and $\dub^1\in L^2(\Omega; H^1_{\#}(\Om_s^\veps)^d)$ such that following convergences hold
	\begin{align}
	\Tuf{\dwbe} 	\rightharpoonup &\dwb^0&\qquad\textrm{w. in }L^2(\Omega\times Y_f),\nonumber\\
	\veps\Tuf{\nabla\dwbe} 	\rightharpoonup &\nabla_y\dwb^0&\qquad\textrm{w. in }L^2(\Omega\times Y_f),\nonumber\\
	\Tuf{\Pe} 	\rightarrow & P^0&\qquad \textrm{ s. in }L^2(\Om),\nonumber\\
	\Tuf{\nabla\Pe} 	\rightharpoonup & \nabla_xP^0+\nabla_yP^1&\qquad \textrm{w. in }L^2(\Omega\times Y_f),\nonumber\\
	\Tuf{\left\{\dPhije\right\}} 	\rightarrow &\left\{\dPhi^0_\valb\right\} &\qquad\textrm{ s. in }L^2(\Om),\nonumber\\
	\Tuf{\left\{\nabla\dPhije\right\}} 	\rightharpoonup &\left\{\nabla_x\dPhi^0_\valb+\nabla_y\dPhi^1_\valb\right\}&\qquad\textrm{w. in }L^2(\Omega\times Y_f),\label{converg_start}\\
	\Tuf{\dPsie}\rightharpoonup&\dPsio&\qquad\textrm{w. in }L^2(\Omega\times Y_f),\nonumber\\
	\Tuf{\veps\nabla\dPsie}\rightharpoonup&\nabla_y\dPsio &\qquad\textrm{w. in }L^2(\Omega\times Y_f),\nonumber\\
	\Tuf{\dube}\rightharpoonup&\dub^0&\qquad\textrm{w. in }L^2(\Omega\times Y_s),\nonumber\\
	\Tuf{\nabla\dube}\rightharpoonup&\nablax\dub^0+\nablay\dub^1&\qquad\textrm{w. in }L^2(\Omega\times Y_s),\nonumber\\
	\Tuf{\sigmabf^\veps_f}\rightharpoonup&\sigmabf^1_f&\qquad\textrm{w. in }L^2(\Omega\times Y_f),\nonumber
	\end{align}
 for $\valbeq$ and where
 
As the consequence of the convergences \Eq{converg_start}, truncated asymptotic expansions of the unfolded unknown fields can be introduced which satisfy the same convergence result. These constitute the recovery sequences in subdomains of $\Om_f^\veps$ and $\Om_s^\veps$ (the respective characteristic functions are $\chi_f^\veps$ and $\chi_s^\veps$) which read

 \begin{eqnarray}
 \Tuf{\chi_f^\veps(x)\dwbe(x)}&\approx&\dwb^0(x,y)+\veps\dwb^1(x,y)+\mathcal{O}(\veps^2),\nonumber\\
 \Tuf{\chi_f^\veps(x)\Pe(x)}&\approx& P^0(x)+\veps P^1(x,y)+\mathcal{O}(\veps^2), \nonumber\\
 \Tuf{\chi_f^\veps(x)\Psie(x)}&\approx&\Psi^0(x)+\veps\Psi^1(x,y)+\mathcal{O}(\veps^2), \nonumber\\
 \Tuf{\chi_f^\veps(x)\dPhije(x)}&\approx&\dPhi^0_\valb(x)+\veps\dPhij^1(x,y)+\mathcal{O}(\veps^2), \valbeq, \label{eq_recseq_start}\\
 \Tuf{\chi_f^\veps(x)\dPsie(x)}&\approx&\dPsi^0(x,y)+\veps\dPsi^1(x,y)+\mathcal{O}(\veps^2),\nonumber\\
 \Tuf{\chi_s^\veps(x)\dube(x)}&\approx&\dub^0(x)+\veps\dub^1(x,y)+\mathcal{O}(\veps^2).\nonumber
 \end{eqnarray}

 Limit functions, $(\dwb^0,P^0,\{\dPhi^0_\valb, \dPhi^1_\valb\},\dPsio(x,y),\dub^0,\dub^1)$ are solutions of the corresponding two-scale limit problems. These are not presented in this paper; 
 we only introduce the multiplicative splits of the two-scale functions which allow us to establish local autonomous problems for characteristic responses.

 \subsection{Scale separation formulas}
 The local problems, relevant to the microscopic scale, can be derived from the limit problem, usually by letting all the components of the test functions, which are not relevant to the microscopic scale.
 Thanks to the linearization of the problem, it is possible to determine scale separation formulae as a linear combination of macroscopic fluxes and corrector base functions, otherwise called the characteristic responses. Two different macroscopic fluxes in the limit two-scale problem are recognized, namely $(\nablax\dPhij^0+\Eb^*)$ and $(\nablax P^0+\fb^*)$.  Therefore, we introduce scale decomposition formulae of the limits $\dwb^0,P^1,\dPhij^1,\valbeq$, which read
 \begin{equation}\label{eq-split1}
   \begin{split}
 \dwb^0(x,y)= \sum\limits_{k=1}^{d}\left(-\ombf^{0,k}(y)\left(\frac{\pd P^0}{\pd x_k} +f^*_k\right)(x)+ \sum\limits_{\valalim}^{2}\ombf^{\vala,k}(y)\left(\frac{\pd \dPhi_\vala^0}{\pd x_k} +E^*_k\right)(x)\right),\\
 P^1(x,y)= \sum\limits_{k=1}^{d}\left(-\pi^{0,k}(y)\left(\frac{\pd P^0}{\pd x_k} +f^*_k\right)(x)+ \sum\limits_{\valalim}^{2}\pi^{\vala,k}(y)\left(\frac{\pd \dPhi_\vala^0}{\pd x_k} +E^*_k\right)(x)\right),\\
  \dPhij^1(x,y)= \sum\limits_{k=1}^{d}\left(-\theta^{0,k}(y)\left(\frac{\pd P^0}{\pd x_k} +f^*_k\right)(x)+ \sum\limits_{\valalim}^{2}\theta^{\vala,k}(y)\left(\frac{\pd \dPhi_\vala^0}{\pd x_k} +E^*_k\right)(x)\right),
   \end{split}
 \end{equation}
 where two families of corrector base functions $(\ombf^{0,k}, \pi^{0,k},\theta_\valb^{0,k})$ and $(\ombf^{\vala,k}, \pi^{\vala,k},\theta_\valb^{\vala,k}),\valbeq,$ were introduced, indexed by $k\in\{1,\dots,d\}$; note $d=3$ is the spatial dimension of the problem.

 The scale separation formulae for the potential and the displacements attain the following forms:
 \begin{equation}\label{eq-split2}
   \begin{split}
 \dPsio(x,y)&=  \sum\limits_{\valblim}^{2}\varpi^\valb(y)\left(\dPhi_\valb^0(x) +\Psieh(x)\right),\\
 \dub^1(x,y)&=  \sum\limits_{i,\valblim}^{d}\wb^{ij}(y)e_{ij}\left(\dub^0(x)\right)+P^0(x)\wb^P(y)+\\
 &+\sum\limits_{\valblim}^{N}\wb^\valb(y)\zj\left(\dPhi_\valb^0(x)  +\Psieh(x)\right),
   \end{split}
 \end{equation}
  where $\wb^{ij},\wb^P$ and $\wb^{\valb}$ are corrector base functions.

 All corrector functions involved in \Eq{eq-split1} and \Eq{eq-split2} are obtained as the solution of three local cell problems given in the next section.

 \subsection{Cell problems}
 Corrector basis functions introduced in \Eq{eq-split1}-\Eq{eq-split2} satisfy the local autonomous  problems (cell problems) defined in subdomains of the representative periodic cell $Y$, therefore.
By virtue of the linearization which decomposes the problem \Eq{eq_lin_start1}-\Eq{eq_lin_stop} in three subproblems, the cell problems are decoupled, thus, the following three groups are distinguished:
 
 \begin{itemize}
 	\item Group 1: two cell problems related to the electrokinetic system with responses $(\ombf^{0,k}, \pi^{0,k},\theta_\valb^{0,k})$ and $(\ombf^{\alpha,k}, \pi^{\alpha,k},\theta_\valb^{\alpha,k}),\alpha,\beta = 1,2$, $k = 1,\dots,d$.
 	\item Group 2: one cell problem related to the electrostatic potential in the EDL with solution $\varpi^\alpha,\valaeq$.
 	\item Group 3: three cell problems related to the  poroelasticity with solutions $,\wb^{ij},\wb^P$ and $\wb^{i}$
 	\end{itemize}
 
 \par While denoting $\mathbf{e}^k$ canonical basis of $\RR^d$, $k=1,...,d$ and $\delta_{ij}$ the Kronecker symbol, all the six cell problems are introduced below.

 \subsubsection{Cell problems: Group 1}
\par The first autonomous cell problem of this group is related to the macroscopic pressure gradient:
 Find $(\ombf^{0,k}, \pi^{0,k},\theta_\valb^{0,k})\in \Hb^1_{\#0}(Y_f)$,  $k=1,2,3,\ \valbeq$:
  \begin{equation}
 \begin{split}\label{eq_lp_el0_1}
 \intYf\nabla\ombf^{0,k}(y):\nabla\testw \dV + \intYf\pi^{0,k}(y)\nabla\cdot\testw\dV&=\intYf\phi_k + \sum\limits_{\valblim}^{2}\intYf \zj\cjo(y)\testw\cdot\nabla\theta_\valb^{0,k} \dV,\\
 \intYf\testp \nabla\cdot\ombf^{0,k}(y)\dV&=0,\\
 \intYf \cjo(y)\nabla\testphi\nabla\theta_\valb^{0,k}\dV+\intYf \cjo(y)\Pecj \zj^{-1}  \testphi\nabla&\cdot\ombf^{0,k}(y)\dV=0,
 \end{split}
   \end{equation}
 for all test functions $\testw\in \Hb^1_{\#0}(Y_f)$, $\testp \in L^2(Y_f)$, $\testphi\in H^1_{\#}(Y_f)^N$.
 \par The second autonomous cell problem, corresponding to the macroscopic diffusive flux, for each species $\valaeq$ reads:
 Find $(\ombf^{\vala,k}, \pi^{\vala,k},\theta_\valb^{\vala,k})\in \Hb^1_{\#0}(Y_f)$, $k=1,2,3,\ \valbeq$:
 \begin{equation}
 \begin{split}\label{eq_lp_eli_1}
 \intYf\nabla\ombf^{\vala,k}(y):\nabla\testw \dV + \intYf\pi^{\vala,k}(y)\nabla\cdot\testw\dV&=\sum\limits_{\valblim}^{2}\intYf \zj\cjo(y)(\delta_{\vala\valb}\mathbf{e}^k+\nabla\theta_\valb^{\vala,k})\cdot\testw \dV,\\
 \intYf\testp \nabla\cdot\ombf^{\vala,k}(y)\dV&=0,\\
 \intYf \cjo^0(y)\nabla\testphi\nabla\theta_\valb^{\vala,k}\dV+\intYf \cjo(y)\Pecj \zj^{-1}  \testphi\nabla\cdot&\ombf^{\vala,k}(y)\dV=-\intYf \cjo(y)\testphi\nabla\cdot(\delta_{\vala\valb}\mathbf{e}^k)\dV,
 \end{split}
  \end{equation}
 for all test functions $\testw\in \Hb^1_{\#0}(Y_f), \testp \in L^2(Y_f), \testphi\in H^1_{\#}(Y_f)^N$.
 
 \subsubsection{Cell problems: Group 2} 
 \par The cell problem associated with the macroscopic ionic potential, for each species $\valaeq$ reads:
 Find corrector base functions $\varpi^\vala\in H^1_{\#}(Y_f),\valaeq$, such that

 \begin{equation}\label{eq_lp_pot}
 \intYf \nabla\varpi^\vala\cdot\nabla\testphi\dV+\betapar\intYf\sum\limits_{\valblim}^{2}\left(\zj^2\cjo(y)\right)\varpi^\vala\testphi\dV=-\intYf \betapar \zi^2 \cio(y)\testphi\dV,
 \end{equation}
 for all test functions $\testphi\in H^1_{\#}(Y_f)$.
 
 \subsubsection{Cell problems: Group 3}
\par The following three cell problems are relevant to the homogenization of displacement perturbation. One can realize, that the first two cell problems are identical to the ones occurring in derivation of Biot's poroelasticity equation. 
The first cell problem reads: find $\wb^{ij}\in\Hpdb(Y_s)^d, \intYi{s} \wb^{ij}\dV=0$ such that
\begin{equation}\label{eq_lp_disp_1}
\intYi{s} \Ab^*\nabla\wb^{ij}:\nabla\testu \dV+\intYi{s} \Ab^*\nabla\Pibf^{ij}:\nabla\testu \dV=0,
\end{equation}
for any test function $\testu\in\Hpdb(Y_s)^d$. Symbol $\Pibf^{ij}$ denotes the so-called transformation vectors $\Pibf^{ij}=(\Pi^{ij}_k), i,j,k=1,\dots,d$, which enable to establish local displacements defined in $Y$ generated by affine transformation of the macroscopic strains $\eb_x(\dub^0)$ defined in $\Om$; it holds that $\eb_y(\Pibf^{ij}e_{ij}^x(\dub^0) = \eb_x(\dub^0)$, where 
\begin{equation}
\Pi^{ij}_k=y_j\delta_{ik}.
\end{equation}
\par The second cell problem reads: find $\wb\in\Hpdb(Y_s), \intYi{s} \wb^P \dV=0$ such that
\begin{equation}\label{eq_lp_disp_2}
\intYi{s}\Ab^*\nabla\wb^P:\nabla\testu \dV=-\intYD{\Gamma_Y}\testu\cdot\nb \dSy,
\end{equation}
for any test function $\testu\in\Hpdb(Y_s)^d$.
\par Finally, the third cell problem connecting displacement perturbation and ionic potentials is needed. It reads: Find $\wb^i\in\Hpdb(Y_s)^d, \intYi{s}\wb^\vala\dV=0$ such that
\begin{align}\label{eq_lp_disp_3}
\intYi{s} \Ab^*\nabla\wb^\vala:&\nabla\testu \dV=\nonumber\\
=\intYD{\Gamma_Y}\testu\cdot&\left(\cjo\Ib+\betaparinv\left(\nabla_y\Psieq\otimes\nabla_y\varpi^\vala+\nabla_y\varpi^\vala\otimes\nabla_y\Psieq-\nabla_y\Psieq\cdot\nabla_y\varpi^\vala\Ib\right)\right)\cdot\nb \dS y,
\end{align}
for any test function $\phibf\in\Hpdb(Y_s)^d$.

\subsection{Macroscopic model}
By virtue of the homogenization method, the limit two-scale equations arising from \Eq{eq_weak_elkin1}-\Eq{eq_weak_disp} involve cell integrals of the two-scale functions which can be expressed in terms of the corrector basis functions. Below we list expressions of the homogenized coefficients which constitute the effective material parameters of the upscaled porous medium, \cite{allaire2013asymptotic}.

The first group of the corrector functions define the following homogenized coefficients
\begin{align}\label{eq_ec_el_1}
\mathcal{J}^\vala_{lk}&=\intYi f\ombf^{\vala,k}(y)\cdot\mathbf{e}^l\dV,\\
\mathcal{K}_{lk}&=\intYi f\ombf^{0,k}(y)\cdot\mathbf{e}^l\dV,\\
\mathcal{D}^{\vala\valb}_{lk}&=\intYi f\left(\ombf^{\vala,k}(y)+\frac{\zj}{\Pecj}\left(\mathbf{e}^k\delta_{\vala\valb}+\nabla_y\theta_\valb^{\vala,k}(y)\right)\right)\cdot\mathbf{e}^l\dV,\\
\mathcal{L}^{\vala}_{lk}&=\intYi f\left(\ombf^{0,k}(y)+\frac{\zi}{\Peci}\nabla_y\theta_\vala^{0,k}(y)\right)\cdot\mathbf{e}^l\dV, \label{eq_ec_el_4}
\end{align}
whereby $\KK=(\mathcal{K}_{lk})$ is the  permeability tensor, $\DD^{\vala\valb}=(\mathcal{D}^{\vala\valb}_{lk})$ are diffusivity tensors; in particular $DD^{\vala\valb}$ describes diffusion of species $\alpha$ due to the streaming potential gradient of the species $\beta$. Tensors $\JJ^\vala=(\mathcal{J}^{\vala}_{lk})$ is related to the flow driven by electric fields
and $\LL^\vala=(\mathcal{L}^{\vala}_{lk})$ , also known as the coupling tensor, expresses the diffusivity of species $\alpha$ due to the global pressure gradient.

The second and the third  group of the corrector functions constitute poroelastic coefficients modified by the presence of the streaming potentials and the external electric field,
\begin{align}\label{eq_ec_disp_1}
\mathcal{A}_{ijkl}&=\intYi{s} \Ab^*\nabla(\wb^{ij}+\Pibf^{ij}):\nabla(\wb^{kl}+\Pibf^{kl})\dV,\\
\mathcal{B}_{ij}&=-\intYi{s} \Ab^*\nabla(\wb^P:\nabla\Pibf^{ij})\dV,\\
\mathcal{C}^\vala_{ij}&=\intYi{s} \Ab^*\nabla(\wb^\vala)\dV+\sum\limits_{\valb=1}^{2}\zj\Ib\intYf \cj^0(y)\left(\varpi^\valb(y)+\delta_{\vala\valb}\right)\dV+\nonumber\\
&+\intYi{f}\betaparinv\left(\nabla_y\Psieq\otimes\nabla_y\varpi^\vala+\nabla_y\varpi^\vala\otimes\nabla_y\Psieq-\nabla_y\Psieq\cdot\nabla_y\varpi^\vala\Ib\right)\dV.\label{eq_ec_disp_3}
\end{align}
Above the tensor $\AAA=(\mathcal{A}_{ijkl})$ is the fourth-order positive definite effective elasticity tensor of drained skeleton,  $\BB^H=(\mathcal{B}_{ij})$ is the Biot's coupling tensor related to the pressure, while $\CC^\vala=(\mathcal{C}^\vala_{kl})$ is the tensor related to ionic potentials.  These effective coefficients are sometimes referred to as the Biot poroelasticity coefficients. For convenience we may introduce coefficient
\begin{equation}
\hat{\BB}=|Y_f|\mathbf{I}+\BB.
\end{equation}

\paragraph{Macroscopic model}
We present the macroscopic model in its dimensional form, \ie using quantities with physical dimensions. The macroscopic variables obtained as the limits of the oscillating solutions of the nondimensionalized problem \Eq{eq_weak_elkin1}-\Eq{eq_weak_disp} can be presented by their  dimensional macroscopic counterparts -- these variables will be denoted by superscript $\sqcup^{\mathrm{eff}}$. Dimensional form of the macroscopic problem reads: Find $({\Peff, \Phieff, \ueff}) \in (L^2(\Om)\times H^1(\Om)^d)$, such that
	\begin{align}\label{eq_macro_elkin1-dimless}
	-\frac{l^2ec_c}{\eta_f}\sum\limits_{\valblim}^{2}\left(\int\limits_{\Om} \JJ^\valb  \nabla_x\Phijeff\nabla_xq \dV + \int\limits_{\Om}\JJ^\valb q\nabla_x\cdot\mathbf{E}\dV \right)+&\int\limits_{\Om}\frac{\KK^D}{\eta_f} \nabla_x \Peff\nabla_xq dV=\nonumber\\
	=&\int\limits_{\Om}\frac{\KK^D}{\eta_f} q\nabla_x\cdot\fb \dV, \\
	-\frac{ec_cD_\vala}{k_BT}\sum\limits_{\valblim}^{2} \left(\int\limits_{\Om}\DD^{ij} \nabla_x\Phijeff\nabla_x\testpsi \dV+ \int\limits_{\Om}\DD^{ij}\varphi\nabla_x\cdot\mathbf{E} \right)\dV+&\int\limits_{\Om}\frac{D_\vala\LL^\vala}{k_BT} \nabla_x \Peff\nabla_x\testpsi \dV=\nonumber\\
	=&\int\limits_{\Om}\frac{D_\vala\LL^\vala}{k_BT} \varphi\nabla_x\cdot \fb \dV, \label{eq_macro_elkin2-dimless}\\
	\int\limits_{\Om}\AAA^Ge_x(\ueff):e_x(\mathbf{\vb})\dV -\int\limits_{\Om}\hat{\BB} \Peff:e_x(\mathbf{\vb}) \dV - ec_c\sum\limits_{\valblim}^{2} \int\limits_{\Om}\CC^\valb&\Phijeff:e_x(\mathbf{\vb}) \dV =\nonumber\\ \label{eq_macro_elast-dimless}
	= \int\limits_{\Om}\fb\cdot\vb \dV+\sum\limits_{\valblim}^{2}&\int\limits_{\Om}\CC^\valb\nabla_x\Psi^{\rm{ext}}\cdot\vb \dV ,
	\end{align}
for all test functions $q\in L^2(\Om), \varphi\in L^2(\Om)$ and $\vb\in H^1(\Om)^d$ and where dimensionalized permeability tensor  is denoted by $\KK^D=l^2\KK$ and  $\AAA^G=\Lambda\AAA$ is Gassmann elasticity tensor. The equation \Eq{eq_macro_elast-dimless}  is  so-called extended Biot equation and it is weakly coupled to the electrokinetic system \Eq{eq_macro_elkin1-dimless} and \Eq{eq_macro_elkin2-dimless} through coefficients $\CC^\vala$. 

\par As reported in the next section, the weak formulation \Eq{eq_macro_elkin1-dimless}-\Eq{eq_macro_elast-dimless} is discretized using the finite elements to obtain the numerical solutions. For the sake of completeness, we also introduce the macroscopic homogenized model in its differential form:

\begin{align}\label{eq_macro_elkin_diff11}
-\nabla_x\cdot\left(\AAA^Ge_x(\ueff)\right) -\nabla_x\cdot\left((|Y_f|\mathbf{I}+\BB) \Peff- ec_c\sum\limits_{\valblim}^{2} \CC^\valb(\Phijeff +\Psi^{\rm{ext}})\right) &=\fb \inOm,  \\
\nabla_x\cdot\left(\frac{l^2 e c_c}{\eta_f}\sum\limits_{\valblim}^{2} \JJ_\valb  (\nabla_x\Phijeff+\mathbf{E})-\frac{\KK^D}{\eta_f} (\nabla_x \Peff -\fb) \right) &=0 \inOm , \label{eq_macro_dif_2}\\
\nabla_x\cdot\left(\frac{ec_cD_\vala}{k_BT}\sum\limits_{\valblim}^{2} \DD_{ij}\left( \nabla_x\Phijeff+\mathbf{E} \right)-\frac{D_\vala\LL_\vala}{k_BT} (\nabla_x \Peff-\fb)\right)
&= 0 \inOm, \label{eq_macro_elkin_diff12}
\end{align}
for  $\valaeq$ and completed by periodic boundary conditions. From the \Eq{eq_macro_dif_2} and \Eq{eq_macro_elkin_diff12} we can distinguish the fluid seepage and the ionic diffusion fluxes,
\begin{align}
\weff&=\frac{l^2ec_c}{\eta_f}\sum\limits_{\valblim}^{2} \JJ_\valb  (\nabla_x\Phijeff+\mathbf{E})-\frac{\KK^D}{\eta_f} (\nabla_x \Peff -\fb)  \inOm, \label{eq_macro_w}\\
\jbi^{\rm{eff}}&=\frac{ec_cD_\vala}{k_BT}\sum\limits_{\valblim}^{2} \DD_{ij}\left( \nabla_x\Phijeff+\mathbf{E} \right)-\frac{D_\vala\LL_\vala}{k_BT} (\nabla_x \Peff-\fb) \inOm \quad \valaeq. \label{eq_macro_j}
\end{align}


\section{Numerical simulation}
\label{sec_numeric}
The aim of this section is to explore and illustrate properties of the homogenized two-scale model described in preceding sections. For this purpose, we present numerical simulation of the  ionic transport through a porous medium occupying a simple-shaped macroscopic domain, with a simple periodic microstructure. We perform a parametric study, which illustrates influence of a varying microstructure on the homogenized material properties. 

	\par The two-scale homogenized model was implemented in \textit{SfePy}, a software for solving problems with coupled partial differential equations (PDEs) in weak forms by means of the finite element method (FEM) for 2D and 3D problems, \cite{sfepy}. \textit{SfePy} is based on the \textit{Python} programming language and its packages \textit{NumPy} and  \textit{SciPy}, \cite{scipy}.
	
	\subsection{Algorithm of numerical implementation}
	\par In the presented examples do not consider volume forces and also disregard effects of an external electric field, thus we put $\fb=\mathbf{0}$ and $\Eb=\mathbf{0}$. The used quantities and parameters are in Tab.~\ref{tab_constants}. All the computations are  performed for the given pore size $l$.
	 
	\par The numerical simulation of the problem can be divided into several steps:
	\begin{enumerate}
		\item Solve the potential distribution in equilibrium $\Psieq$ on cell $Y$ as a solution of \Eq{eq_adim_PB0}.
		\item Compute concentrations $\cjo,\valbeq$ from \Eq{eq_concentration}.
		\item Compute corrector functions $(\ombf^{0,k},\pi^{0,k},\theta_\valb^{0,k})$ and $(\ombf^{\vala,k},\pi^{\vala,k},\theta_\valb^{\vala,k})$, $k=1,\dots,d$, $\valaeq$, $\valbeq$, related to electrokinetic system as a solution of local problems \Eq{eq_lp_el0_1} and \Eq{eq_lp_eli_1}.
		\item Compute effective coefficients relevant to the decoupled electrokinetic system from \Eq{eq_ec_el_1}-\Eq{eq_ec_el_4}.
		\item Compute corrector functions  $\varpi^{\vala}, \valaeq$ related to the potential perturbation as a solution of local problems \Eq{eq_lp_pot}.
		\item Compute corrector functions  $(\wb^{ij}, \wb^P, \wb^\vala)$, $\valaeq$, $i,j=1,\dots,d$, related to the displacement perturbation as a solution of local problems \Eq{eq_lp_disp_1}-\Eq{eq_lp_disp_3}.
		\item Compute effective coefficients relevant to the Biot poroelasticity from \Eq{eq_ec_disp_1}-\Eq{eq_ec_disp_3}.
		\item Compute solution to the macroscopic homogenized system of equations \Eq{eq_macro_elkin1-dimless}-\Eq{eq_macro_elast-dimless}.
	\end{enumerate}
	
	Since the major part of the equations were presented in their dimensionless form, all the homogenization results will be presented in the dimensionless form as well, unless stated otherwise.
	 
	\subsection{Geometrical representation of microstructure}
	\par
\begin{figure}[t]
			\begin{subfigure}{0.5\linewidth}	
				\centering
				\includegraphics[width=\linewidth]{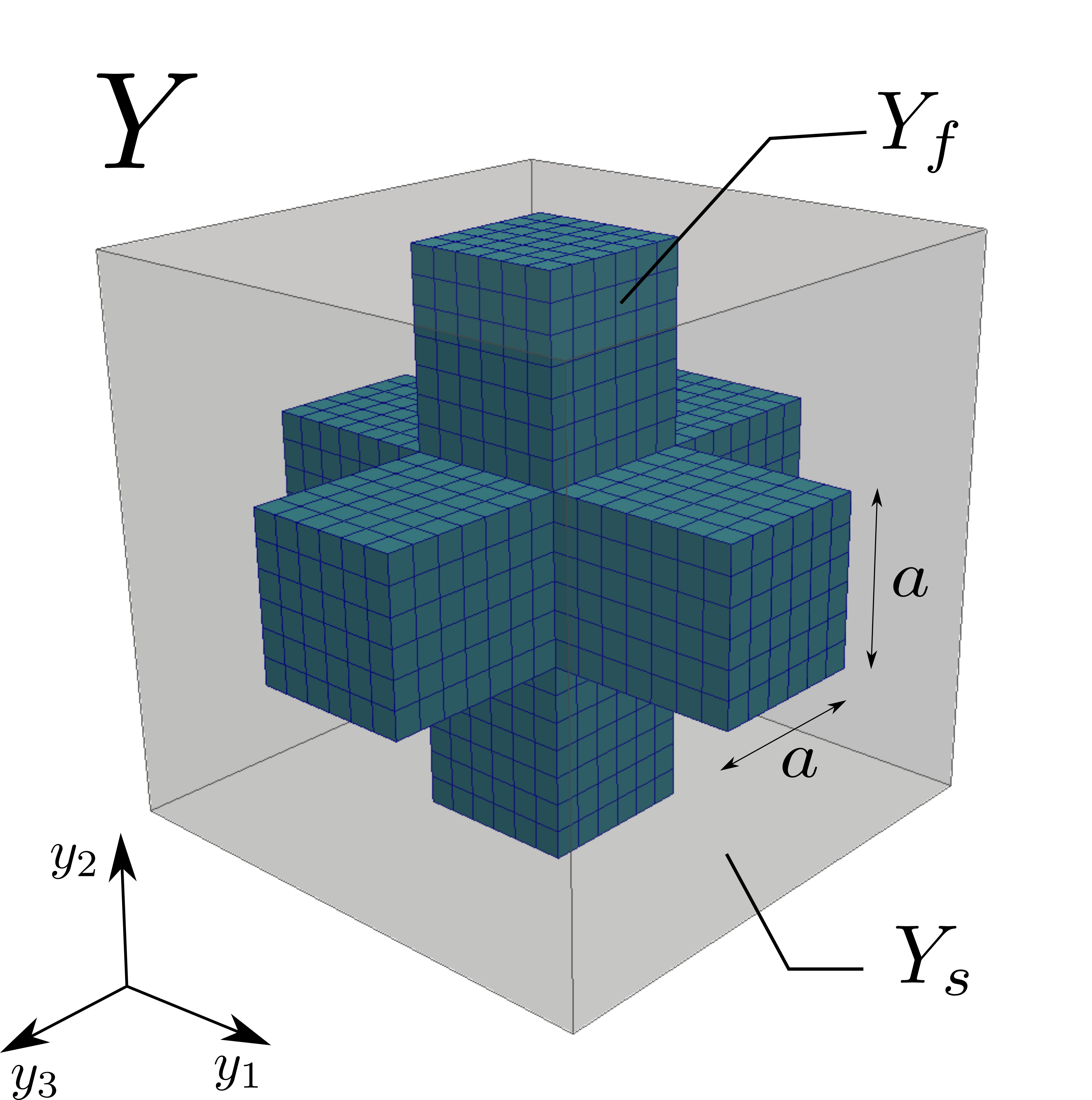}	
				\caption{}	
				\label{cells}
			\end{subfigure}
			\begin{subfigure}{0.5\linewidth}
				\centering
				\includegraphics[width=\linewidth]{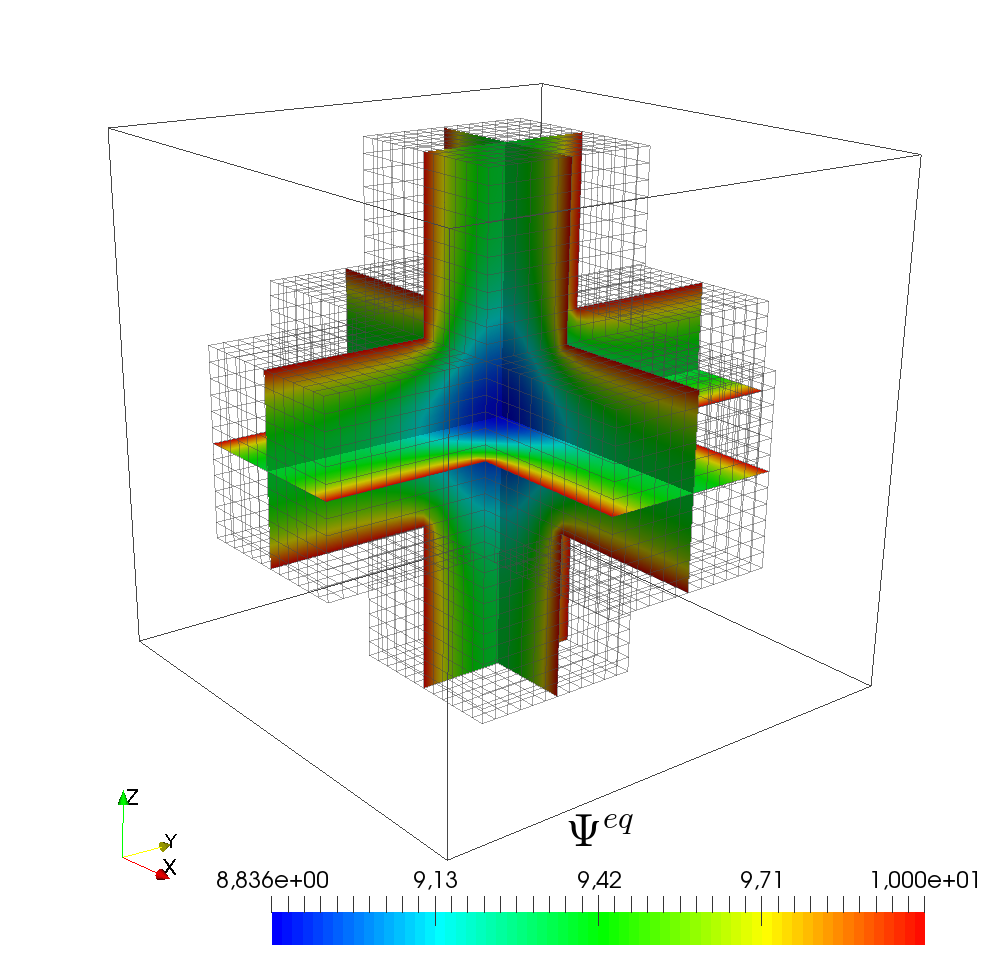}
				\caption{}
				\label{fig_pot0}
			\end{subfigure}
			\caption{Left: Geometry representation of microstructure, parametrization of RPC $Y$; Right: Potential $\Psieq[-]$ distribution on microscale, solution of Poisson-Boltzmann problem in equilibrium.}
			\label{cross}
		\end{figure}

 We aim to study the dependency of the effective coefficients on a change of the microstructure. For this purpose we choose only a simple geometry representation in the form of three interconnected canals aligned with $y_1$-, $y_2$- and $y_3$-directions trough a continuous matrix, see Fig.~\ref{cells}. The cross-section of the canals is a square with size $a$. Changing the parameter $a$ leads to a change in porosity $\hat\phi_f=\frac{|Y_f|}{|Y|}$. 
	\par The mesh representing the cell $Y$ was generated by a mesh generation script, which forms a part of the \textit{SfePy} software. For meshing purposes, the linear hexahedron elements were used.
	 
	\subsection{Semi-discretized macroscopic problem}

	We introduce the semi-discretized form of the macroscopic problem, which can be used in the finite element (FE) model. By $\Pbeff,\Phibeff_\vala \valaeq$ and $\ubeff$ we refer to the column vectors incorporating all degrees of freedom of FE mesh nodes associated with partitioning of the macroscopic domain $\Om$. We need the following approximations of the terms involved in the macroscopic problem \Eq{eq_macro_elkin1-dimless} - \Eq{eq_macro_elast-dimless}:
	\begin{displaymath}
	\begin{array}{lcllcl}
	\qb^T\Kbdisc\Pbeff & \approx & \int_{\Om}\frac{\KK^D}{\eta_f} \nabla_x \Peff\nabla_xq dV, &\qquad \qb^T\Jbdisc^\vala\Phibeff_\vala& \approx &\frac{l^2ec_c}{\eta_f}\int_{\Om} \JJ^\vala \nabla_x\Phijeff\nabla_xq \dV, \\
	\qb^T\fbdisc_J^{\vala} & \approx & \int_{\Om}\JJ^\vala q\nabla_x\cdot\mathbf{E}\dV, &\qquad  \qb^T\fbdisc_K& \approx &\int_{\Om}\frac{\KK^D}{\eta_f} q\nabla_x\cdot\fb \dV, \\
	\sbb^T\Lbdisc^\vala\Pbeff& \approx &\int_{\Om}\frac{D_\vala\LL^\vala}{k_BT} \nabla_x \Peff\nabla_xs \dV,  &\qquad \sbb^T\Dbdisc^{\vala\valb}\Phibeff_\valb& \approx & \frac{ec_cD_\vala}{k_BT}\int_{\Om}\DD^{\vala\valb} \nabla_x\Phijeff\nabla_xs \dV,\\
	\sbb^T\fbdisc_D^{\vala\valb} & \approx & \int_{\Om}\DD^{\vala\valb}s\nabla_x\cdot\mathbf{E} \dV, &\qquad \rb^T\Bbdisc\Pbeff& \approx & \int_{\Om}\hat{\BB} \Peff:e_x(\mathbf{\rb}) \dV,\\
	\rb^T\Cbdisc^\vala\Phibeff_\vala & \approx & ec_c\int_{\Om}\CC^\vala\Phieff:e_x(\mathbf{\rb}) \dV, &\qquad \rb^T\Abdisc\ubeff& \approx & \int_{\Om}\AAA^Ge_x(\ueff):e_x(\mathbf{\rb})\dV,\\
	\rb^T\fbdisc_C^\vala & \approx & \int_{\Om}\CC^\vala\nabla_x\Psi^{\rm{ext}}\cdot\rb \dV, &\qquad \rb^T\fbdisc & \approx & \int_{\Om}\fb\cdot\rb \dV. \end{array}
	\end{displaymath}
	
	\par Using the notations just introduced, we can write the linear macroscopic problem in the matrix form 
	\begin{displaymath}\label{mtx_form}
	\left[\begin{array}{cccc}
	\Kbdisc & -\Jbdisc^1 & -\Jbdisc^2 & \mathbf{0}\\
	\Lbdisc^1 & -\Dbdisc^{11} & -\Dbdisc^{12} & \mathbf{0}\\
	\Lbdisc^2 & -\Dbdisc^{21} & -\Dbdisc^{22} & \mathbf{0}\\
	\Bbdisc &  -\Cbdisc^1 & -\Cbdisc^2 & \Abdisc
	\end{array}\right]
	\left[\begin{array}{c}
	\Pbeff\\
	\Phibeff_1\\
	\Phibeff_2\\
	\ubeff
	\end{array}\right] =
	\left[\begin{array}{c}
	\fbdisc_J^1+\fbdisc_J^2+\fbdisc_K\\
	\fbdisc^{11}_D + \fbdisc^{12}_D\\
	\fbdisc^{21}_D + \fbdisc^{22}_D\\
	\fbdisc_C^1+\fbdisc_C^2+\fbdisc
	\end{array}\right].
	\end{displaymath}  
	From the macroscopic problem \Eq{eq_macro_elkin_diff11}-\Eq{eq_macro_elkin_diff12} is immediately evident, that the electrokinetic system can be solved separately from the problem of poroelasticity, thus the solution $(\Pbeff,\Phibeff_\vala), \valaeq$ is obtained. Than, the macroscopic displacement can be found as
	\begin{displaymath}
	\ubeff=\Abdisc^{-1}\left(- \Bbdisc\Pbeff   +\Cbdisc^1\Phibeff_1 +\Cbdisc^2\Phibeff_2 +\Cbdisc_\Eb^1+\Cbdisc_\Eb^2+\fbdisc \right).
	\end{displaymath}

	\subsubsection{Potential distribution in equilibrium}
	The first step in obtaining effective coefficients is to compute the distribution of potential $\Psieq$  in equilibrium on the microscopic scale. All used electrokinetic quantities can be found in Tab.~\ref{tab_constants}. The distribution of the dimensionless equilibrium potential $\Psieq$ is shown in Fig.~\ref{fig_pot0}.  
	The potential $\Psieq$ has its maximum on the solid-fluid interface, where the surface charge $\Sigma^*$ is prescribed. The potential $\Psieq$ gradually decreases with increasing distance from the interface. This meets our general expectation about Poisson-Boltzmann potential distribution near the solid-fluid interface. The resulting potential distribution $\Psieq$ is needed for computation of concentration $\cjo$ and for subsequential calculations.
	
		\begin{figure}[t]\centering
			\includegraphics[width=\linewidth]{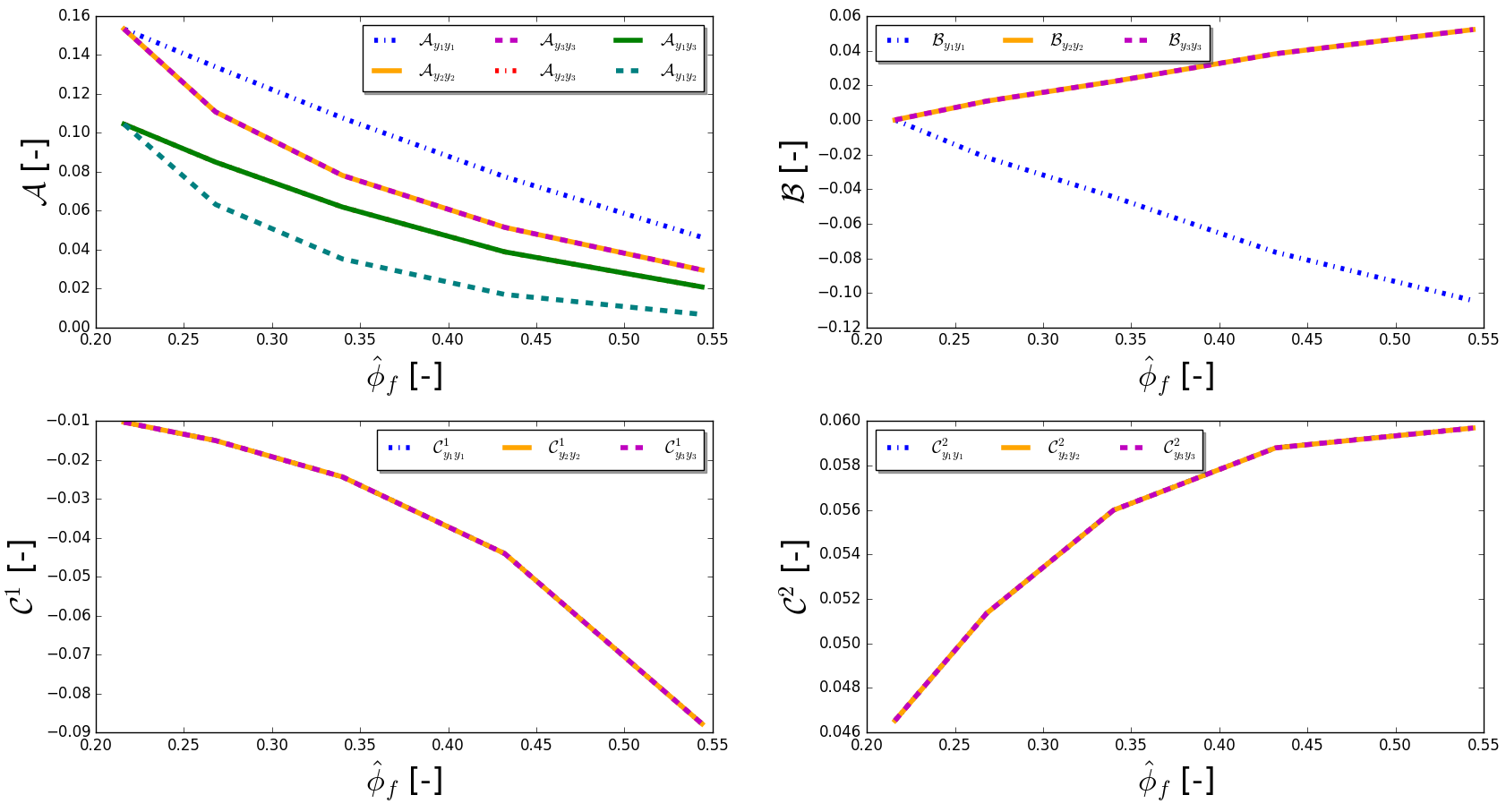}
			\caption{Dependency of components of dimensionless effective elasticity tensor $\AAA$, ionic potential tensors $\CC^1,\CC^2$ and Biot's tensor $\BB$  on porosity $\hat{\phi_f}$.}
			\label{fig_coef_elastic}
		\end{figure}
		\begin{figure}[h]
			\begin{subfigure}{\linewidth}
				\centering
				\includegraphics[width= \linewidth]{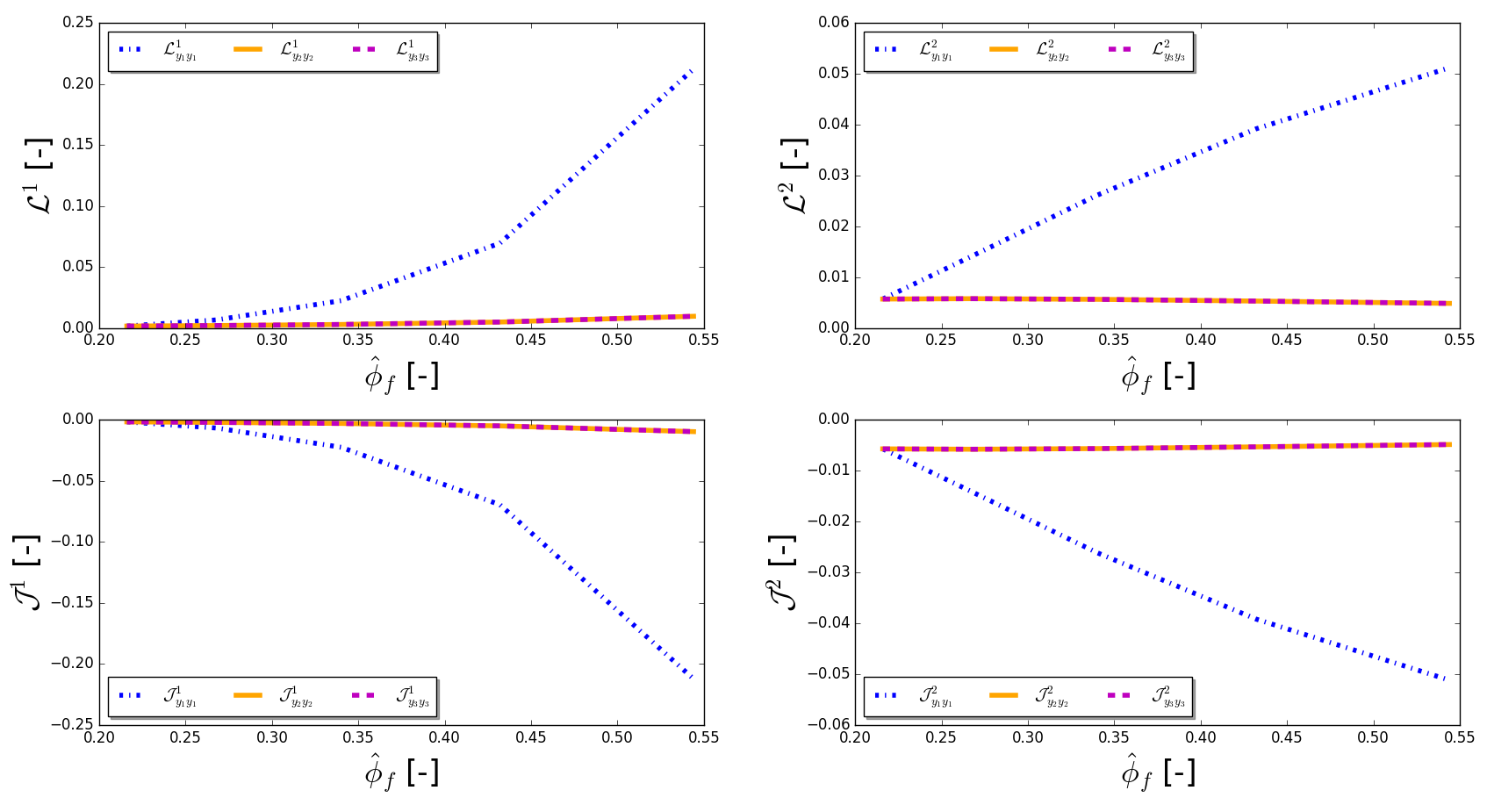}
			\end{subfigure}
			\begin{subfigure}{0.5\linewidth}
				\includegraphics[width=\linewidth]{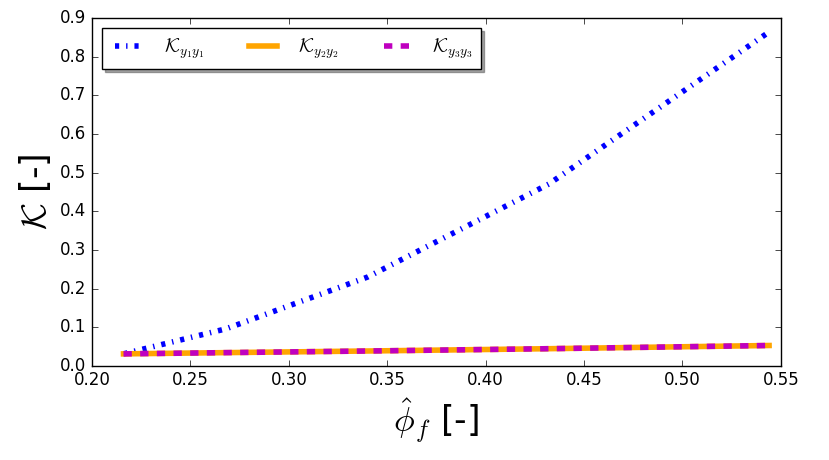}
			\end{subfigure}
			\begin{subfigure}{0.5\linewidth}
				\includegraphics[width=\linewidth]{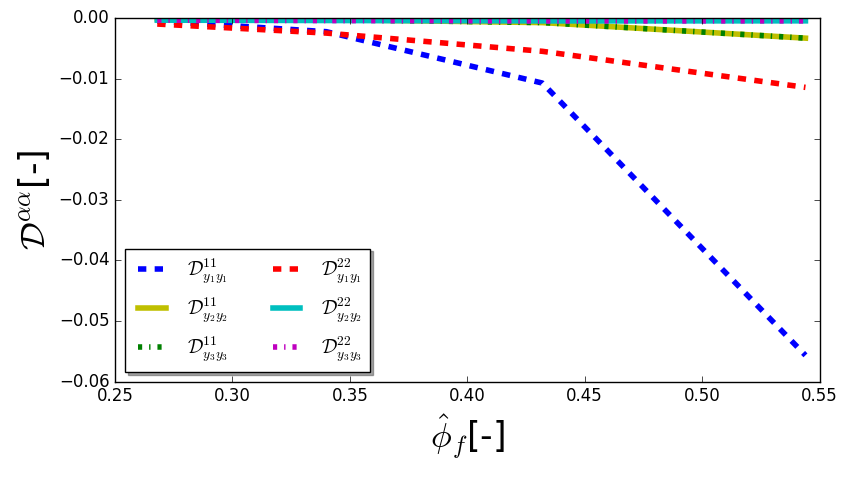}
			\end{subfigure}
			\caption{Dependency of dimensionless effective tensors $\JJ^1,\JJ^2$ relevant to migration-diffusion, coupling tensors $\LL^1,\LL^2$, diffusivity tensors $\DD^{11},\DD^{22}$ and permeability tensor $\KK$ on porosity $\hat{\phi}_f$.}
			\label{fig_coef_elkin}
		\end{figure}
	
	\subsubsection{Influence of varying microstructure on effective tensors}
	\par We study the dependency of effective coefficients on the change in porosity $\hat\phi_f$, caused by variation in $y_1$-direction canal size parameter $a$. These dependencies of the diagonal components of electrokinetic and poroelasticity coefficients (dimensionless) can be seen in Fig.~\ref{fig_coef_elastic} and Fig.~\ref{fig_coef_elkin}. 
	In accordance with the choice of the symmetric microstructure, the components of effective coefficients related to $y_2$- and $y_3$-direction are expected to be equal. 
	
	Fig.~\ref{fig_coef_elastic} shows the dependency of the effective coefficient related to poroelasticity on the porosity of the microstructure. The upper left graph shows components of dimensionless poroelasticity tensor $\AAA$ and its dependency on the microstructure porosity. As expected, all the components of the poroelasticity decrease with the increasing porosity.  In the upper right graph, the Biot coefficient $\BB$ increases with the porosity. Finally, the lower half of the figure shows ionic potential tensors $\CC^1$ and $\CC^2$. The components of $\CC^1$ related to all three direction are equal. This property applies to the components of $\CC^2$ as well. However, the tensor $\CC^1$ decrease  and the tensor $\CC^2$ increase nonlinearly with the increasing porosity.

	Similar nonlinear behavior is obtained for the other electrokinetic tensors as seen in Fig.~\ref{fig_coef_elkin}. We observe the decrease in the components of the migration-diffusion tensors  $\JJ^1$ and $\JJ^2$ related to anions  and cations, respectively.  The components of coupling tensors $\LL^1$  and $\LL^2$ decrease with the increasing porosity.
	
	In the lower part of Fig.~\ref{fig_coef_elkin} we observe increase in permeability $\KK$, as expected. The last part of this figure depicts the decreasing components of diffusivity tensors $\DD^{11}$ and $\DD^{22}$. The diagonal components of $\DD^{12}$ and $\DD^{21}$ are identic with those of $\DD^{22}$.

	\begin{figure}[t]\centering
		\includegraphics[width=\linewidth]{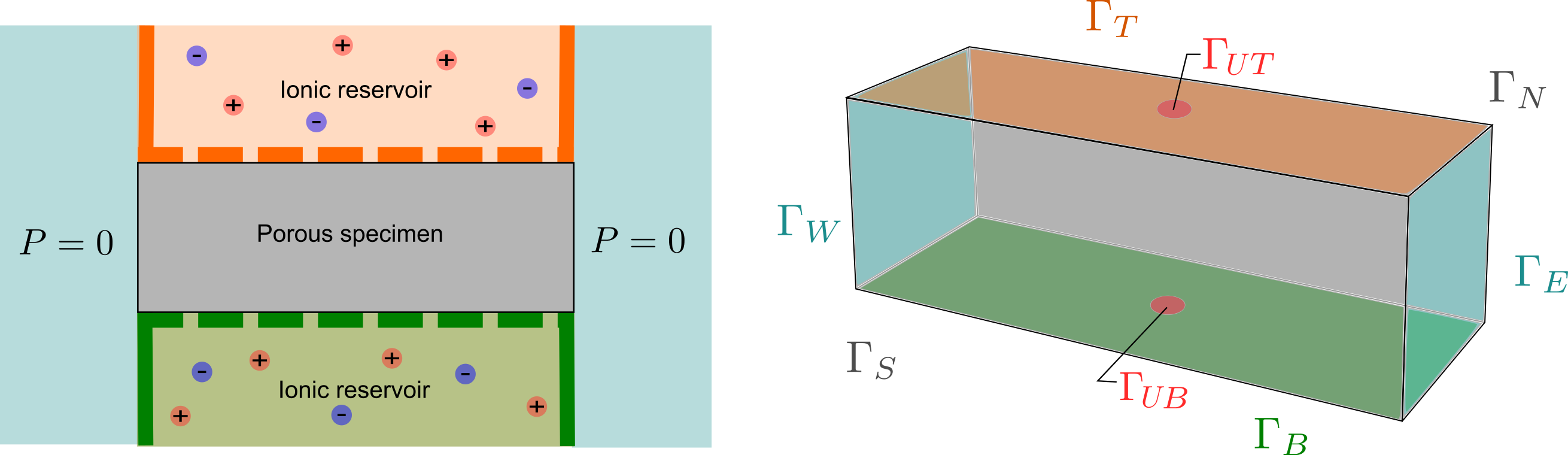}
		\caption{ Simple test geometry and boundary conditions for computation of homogenized macroscopic problem.}
		\label{fig_bc}
	\end{figure}
	

	\subsubsection{Solution of macroscopic problem}

	For the purpose of numerical simulations we propose a simple experiment, see Fig.~\ref{fig_bc}, where a small cuboid specimen occupied by a porous medium is placed between two ionic reservoirs and separated by semipermeable membranes. These membranes enable ionic exchange, but prevent fluid flow.
		
	To describe the boundary conditions, we refer to the faces of the porous specimen by the intuitive notation, such that $\Gamma_E$ stands for the “east side” with the normal vector aligned with $x_1$-axis, whereas the “north side” $\Gamma_N$ has its normal aligned with $x_2$-axis. Then $\Gamma_T$ and $\Gamma_B$ refer to the top and bottom sides, respectively, see Fig.~\ref{fig_bc}. On the ``top'' and ``bottom'' boundaries $\Gamma_{UT}$ and $\Gamma_{UB}$, respectively, the porous specimen is clamped.
	
	This experiment is focused on the observation of the displacement and pressure distribution under the ionic potentials change. To this aim, we propose four macroscopic problems with varying boundary conditions related to the ionic potentials. 
	The homogenized macroscopic problem is given by the system of equations \Eq{eq_macro_elkin1-dimless}-\Eq{eq_macro_elast-dimless} completed by its respective boundary conditions. In what follows we define the two sets of boundary conditions, thus obtaining two boundary problems. The two boundary value problems (BVP) are defined in the following part.
	
        In order to prevent numerical errors, the problem is computed in its dimensionless form. Then, using the dimensionless choices from  Section~\ref{sec:dimless}, we recover dimensional form of respective macroscopic quantities. 
		\begin{figure}[h]		\centering
			\begin{subfigure}{0.49\linewidth}
				\centering
				\includegraphics[width=0.95\linewidth]{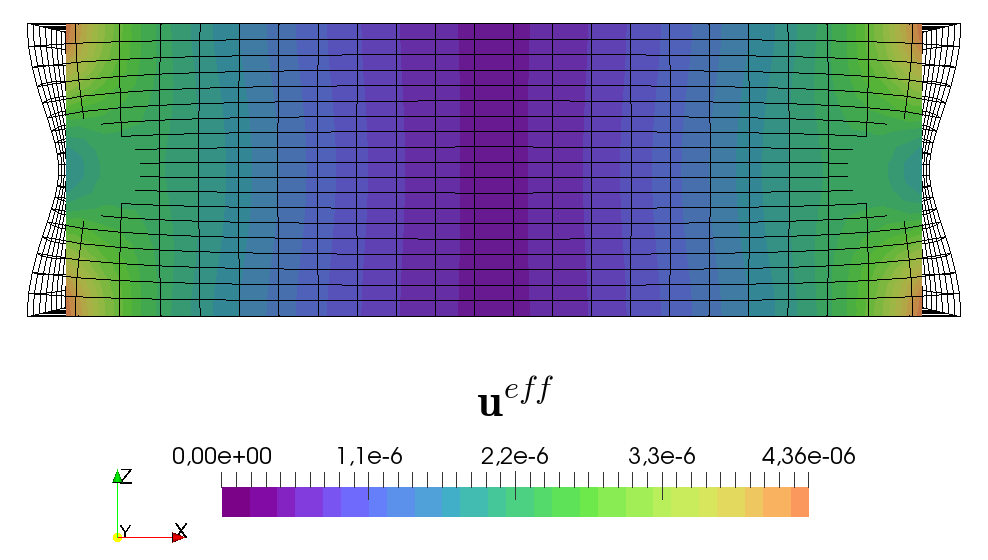}
				\caption{Distribution of displacement $\ub^\textrm{eff}[m]$ in porous specimen.}
			\end{subfigure}
			\begin{subfigure}{0.49\linewidth}	\centering
				\includegraphics[width= 0.95\linewidth]{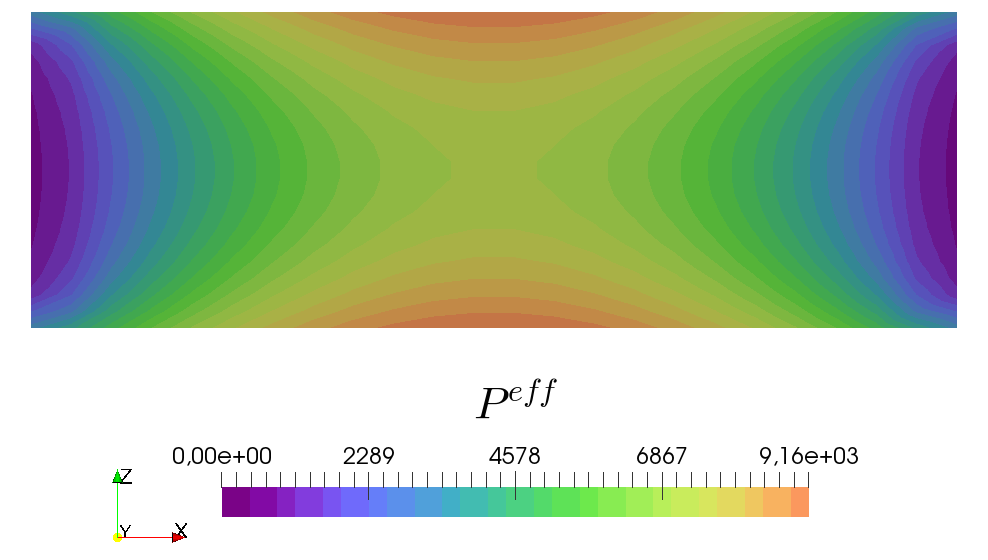}
				\caption{Distribution of  global pressure $P^\textrm{eff}[Pa]$ in porous specimen.}
			\end{subfigure}
			\begin{subfigure}{0.49\linewidth}
				\centering
				\includegraphics[width=0.95\linewidth]{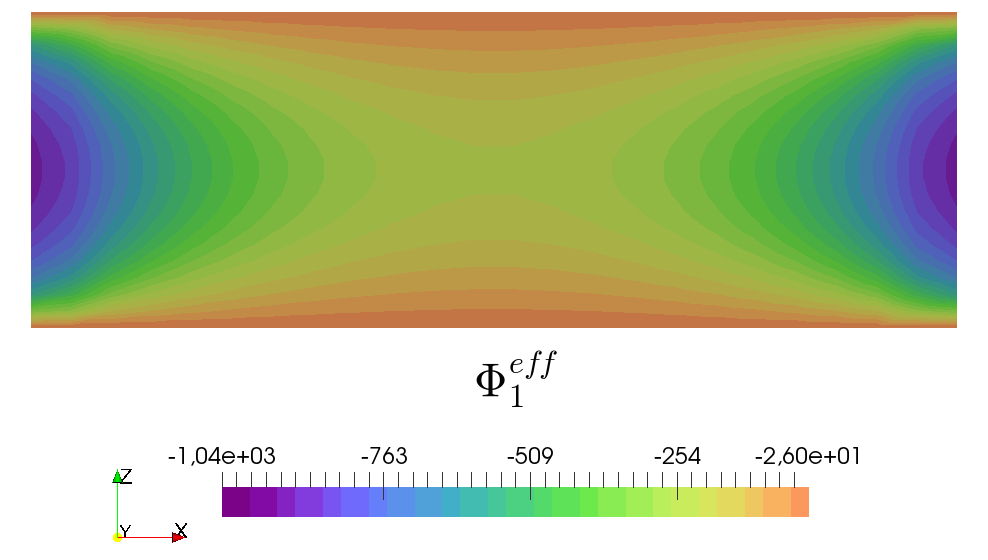}
				\caption{Distribution of  potential $\Phi_1^\textrm{eff}[JC^{-1}]$ in porous specimen.}
			\end{subfigure}
			\begin{subfigure}{0.49\linewidth}	\centering
				\includegraphics[width= 0.95\linewidth]{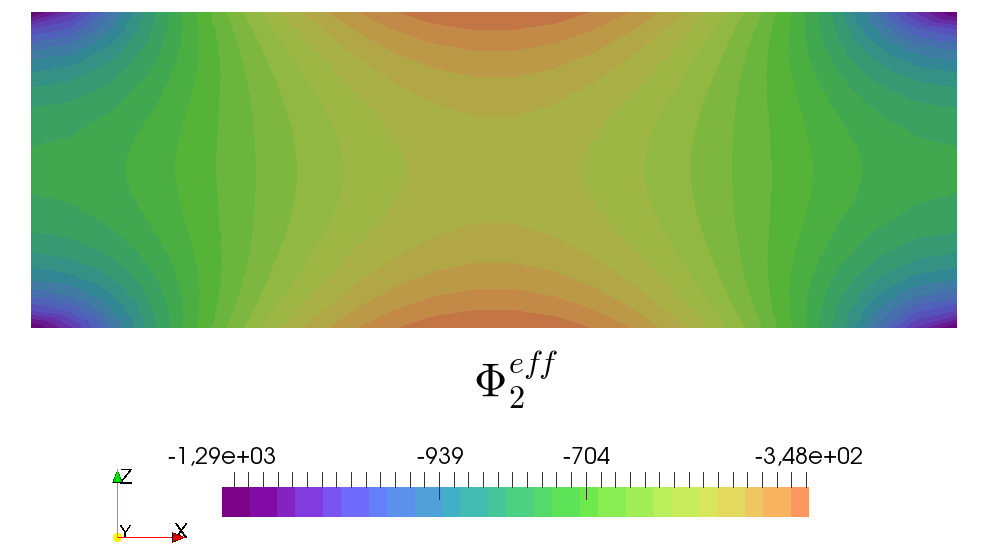}
				\caption{Distribution of  potential $\Phi_2^\textrm{eff}[JC^{-1}]$ in porous specimen.}
			\end{subfigure}
			\caption{Solution of homogenized macroscopic problem, BVP I.}
			\label{fig_macro_sol1}
		\end{figure}
	\subsubsection{Boundary value problems}
	The boundary value problem is defined by \Eq{eq_macro_elkin1-dimless}-\Eq{eq_macro_elast-dimless} and by the following boundary conditions:
	\begin{itemize}
		\item on $\Gamma_{UT}$ and $\Gamma_{UB}$: $\ueff=\mathbf{0}$
		\item on $\Gamma_{W}$ and $\Gamma_{E}$: $\Peff=0$
		\item on $\Gamma_{T}$: $\Phieffone=\bar{\Phi}_1,\quad \JJ^2\nabla\Phiefftwo\cdot\nb=\bar{g}$
		\item on $\Gamma_{B}$: $\Phieffone=\bar{\Phi}_1,\quad \JJ^2\nabla\Phiefftwo\cdot\nb=b\bar{g}$
	\end{itemize}
	The values of boundary conditions are $\bar{\Phi}_1=0.1$ and $\bar{g}=0.001$. By the choice of parameter $b$ we distinguish two BVPs. 
	
	The first boundary problem (BVP I) is defined by the choice $b=1$, so that the boundary conditions are symmetric on boundaries $\Gamma_{T}$ and $\Gamma_{B}$. Therefore, the symmetric  distribution of macroscopic quantities is obtained correspondingly, as seen in Fig.~\ref{fig_macro_sol1}. The deformed shape is visualized by the wire-frame, whereby the displacement field is enlarged by factor $2\cdot 10^5$.  The swelling of the macroscopic body occurs mainly in the region, where $\Phiefftwo$ attains the lowest values.

		\begin{figure}[h]		\centering
			\begin{subfigure}{0.49\linewidth}
				\centering
				\includegraphics[width=0.95\linewidth]{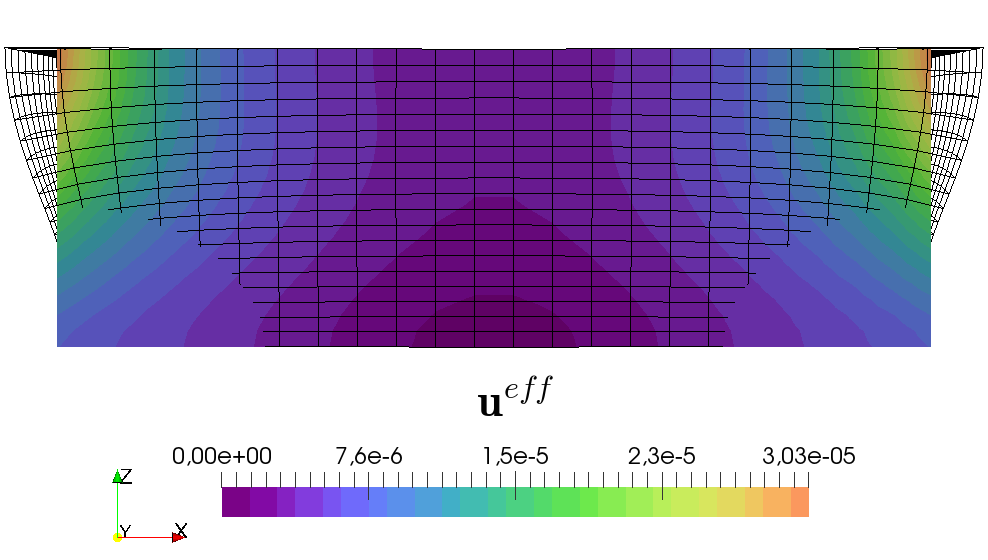}
				\caption{Distribution of dimensionless displacement $\ub^\textrm{eff}[m]$ in porous specimen.}
			\end{subfigure}
			\begin{subfigure}{0.49\linewidth}	\centering
				\includegraphics[width= 0.95\linewidth]{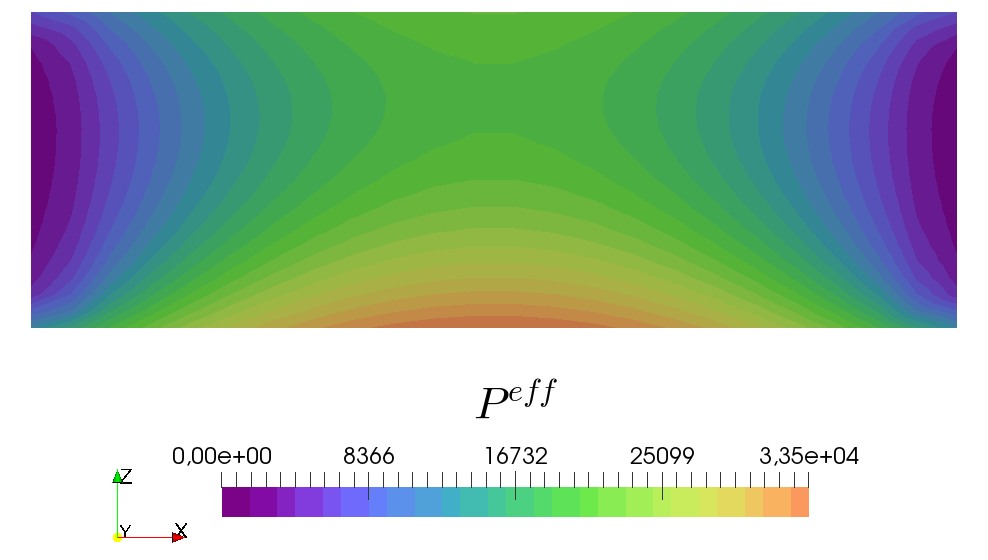}
				\caption{Distribution of dimensionless global pressure $P^\textrm{eff}[Pa]$ in porous specimen.}
			\end{subfigure}
			\begin{subfigure}{0.49\linewidth}
				\centering
				\includegraphics[width=0.95\linewidth]{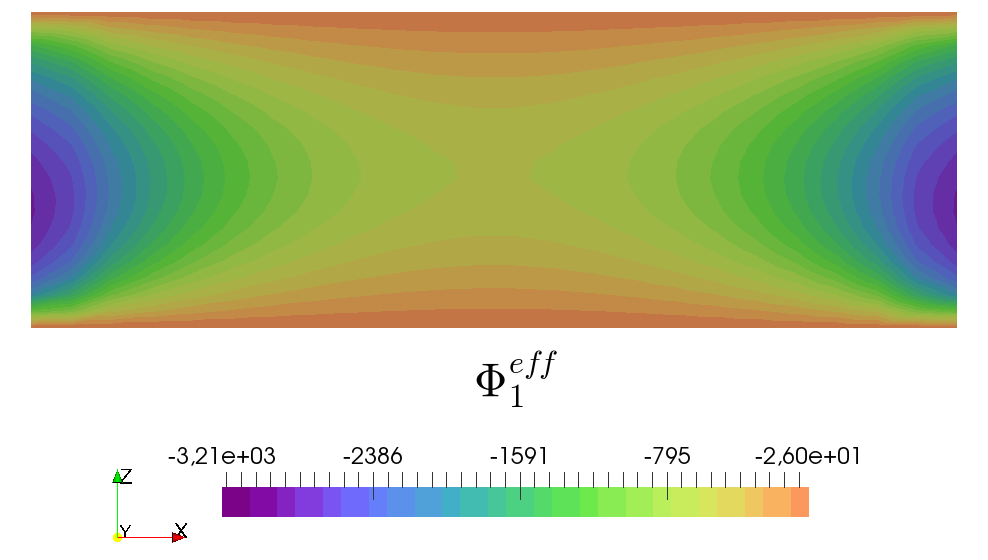}
				\caption{Distribution of dimensionless potential $\Phi_1^\textrm{eff}[JC^{-1}]$ in porous specimen.}
			\end{subfigure}
			\begin{subfigure}{0.49\linewidth}	\centering
				\includegraphics[width= 0.95\linewidth]{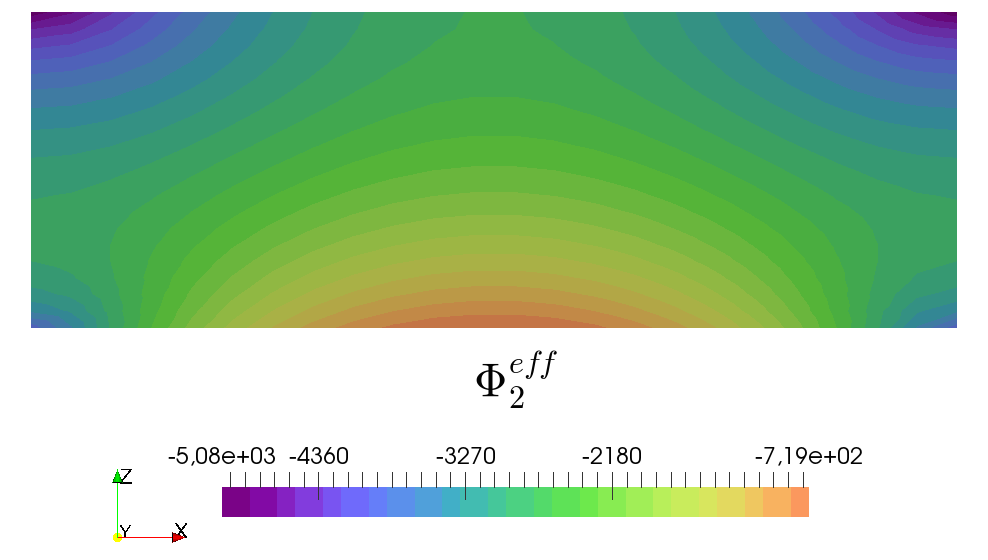}
				\caption{Distribution of dimensionless potential $\Phi_2^\textrm{eff}[JC^{-1}]$ in porous specimen.}
			\end{subfigure}
			\caption{Solution of homogenized macroscopic problem, BVP II.}
			\label{fig_macro_sol3}
		\end{figure}
	By taking $b=5$ we get the second boundary value problem (BVP II). In this case, we increased the influx of $\Phiefftwo$ on the boundary $\Gamma_B$. This leads to the contraction of the porous specimen near this surface, as seen in Fig.~\ref{fig_macro_sol3}. Naturally, the non-symmetrical boundary conditions result  in a non-symmetric distribution of the macroscopic quantities. This effect is most visible on the swelling of the macroscopic body, which tends to react to the distribution of $\Phiefftwo$, being slightly more swelled where $\Phiefftwo$ is slightly lower. The deformed shape is visualized by the wire-frame enlarged by factor $2\cdot 10^4$ only, in this case.

	\subsection{Reconstruction of macroscopic solution on microstructure level}
	One of the most remarkable advantages of the homogenization method is the possibility to reconstruct the solution at the microscopic scale. After computing the global (dimensionless) responses $\left\{\ub^0, P^0, \Phio_\valb\right\}$, it is possible to reconstruct the associated microscopic quantities. This process is also called downscaling in contrast to the upscaling process leading to the macroscopic model. Let us now briefly introduce the reconstruction relations.
	
	\par We consider a given $\veps>0$ corresponding to a real size of the microstructure. This enables us to apply the decomposed forms of the fluctuating two-scale functions in \Eq{eq-split1}-\Eq{eq-split2} $\ub^0, P^0, \Phiio,\valaeq$ defined in domain $\Om$ with the local characteristic responses (corrector basis functions) defined in $Y$. The local microscopic fields are given by the so-called folding mapping (\cite{rohan2017darcy}), such that
	\begin{equation}
	\fold(\hat{x}):(\ub^0, P^0, \Phiio)\rightarrow(\ub^{\mic,\veps},P^{\mic,\veps},\wb^{\mic,\veps},\Phi_\vala^{\mic,\veps})(y),\qquad y\in Y.
	\end{equation}
	The folding operator $\fold$ combines corrector basis functions defined in $Y$ with interpolated macroscopic responses transformed to the zoomed cell $Y$ by the operator  $\Qcalmic$. The operator $\Qcalmic$ is average operator over the recovery cell. 
	
	Using this approach, the microscopic fields $\ub^{\mic,\veps},\wb^{\mic,\veps}, P^{\mic,\veps}, \Phi_\valb^{\mic,\veps},\valbeq$ can be reconstructed as follows:
	
	\begin{equation}
	\begin{split}
	\ub^{\mic,\veps}&= \Qcalmic(\ub^0) +\sum\limits_{i,j=1}^{d}\wb^{ij}\Qcalmic(e_{ij}\left(\ub^0\right))+\Qcalmic(P^0)\wb^P+\sum\limits_{\valblim}^{2}\wb^\valb\zj\Qcalmic\left(\Phi_\valb^0 +\Psieh\right),\label{eq_rec_disp}
	\\
	\wb^{\mic,\veps}&=-\sum\limits_{k=1}^{d}\left(\ombf^{0,k}\Qcalmic\left(\frac{\pd P^0}{\pd x_k} +f^*_k\right)- \sum\limits_{\valalim}^{2}\ombf^{\vala,k}\Qcalmic\left(\frac{\pd \Phiio}{\pd x_k} +E^*_k\right)\right),\\
	P^{\mic,\veps}&=\Qcalmic(P^0)- \sum\limits_{k=1}^{d}\left(\pi^{0,k}\Qcalmic\left(\frac{\pd P^0}{\pd x_k} +f^*_k\right)- \sum\limits_{\valalim}^{2}\pi^{\vala,k}\Qcalmic\left(\frac{\pd \Phi_\vala^0}{\pd x_k} +E^*_k\right)\right),\\
	\dPsi^{\mic,\veps}&= \Qcalmic(\dPsio) +\sum\limits_{\valblim}^{N}\varpi^\valb\Qcalmic\left(\Phi_\valb^0 +\Psieh\right),\\
	\Phi_\valb^{\mic,\veps}&=\Qcalmic(\Phijo)- \sum\limits_{k=1}^{d}\left(\theta^{0,k}\Qcalmic\left(\frac{\pd P^0}{\pd x_k} +f^*_k\right)- \sum\limits_{\valalim}^{2}\theta^{\vala,k}\Qcalmic\left(\frac{\pd \Phi_\vala^0}{\pd x_k} +E^*_k\right)\right).
	\end{split}
	\end{equation}
	
	\begin{figure}[!ht]		\centering
		\begin{subfigure}{0.49\linewidth}
			\centering
			\includegraphics[width=0.95\linewidth]{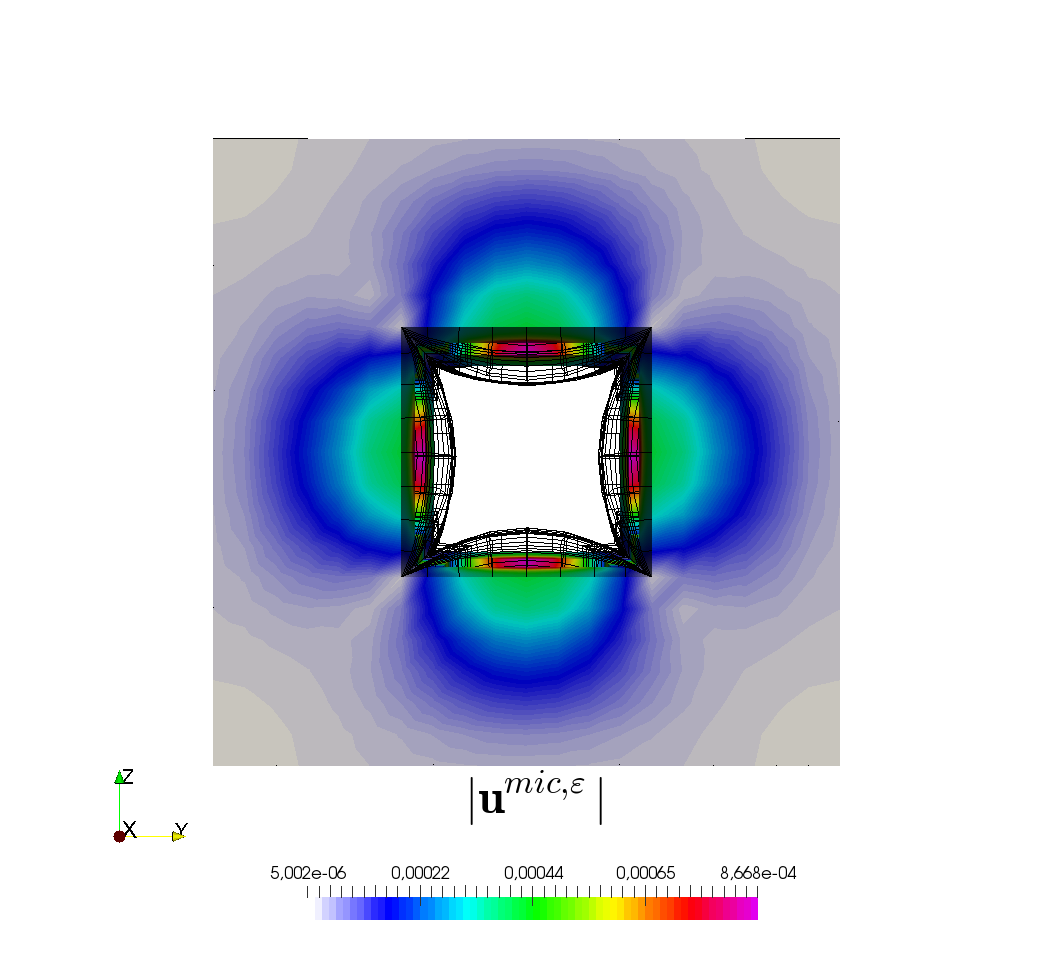}
			\caption{Reconstructed displacement $\ub^{\mic,\veps}$  in 2D view.}
		\end{subfigure}
		\begin{subfigure}{0.49\linewidth}	\centering
			\includegraphics[width= 0.95\linewidth]{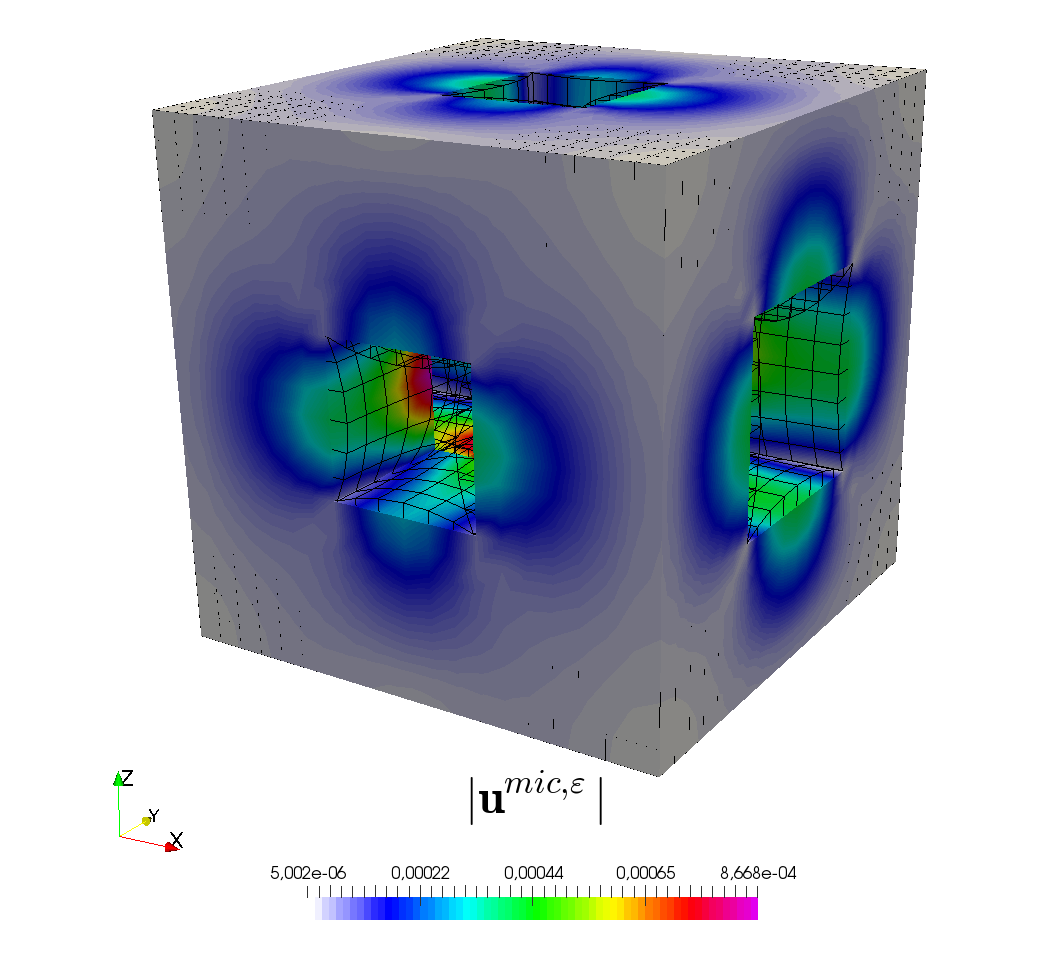}
			\caption{Reconstructed displacement $\ub^{\mic,\veps}$  in 3D view.}
		\end{subfigure}
		\begin{subfigure}{0.49\linewidth}
			\centering
			\includegraphics[width=0.95\linewidth]{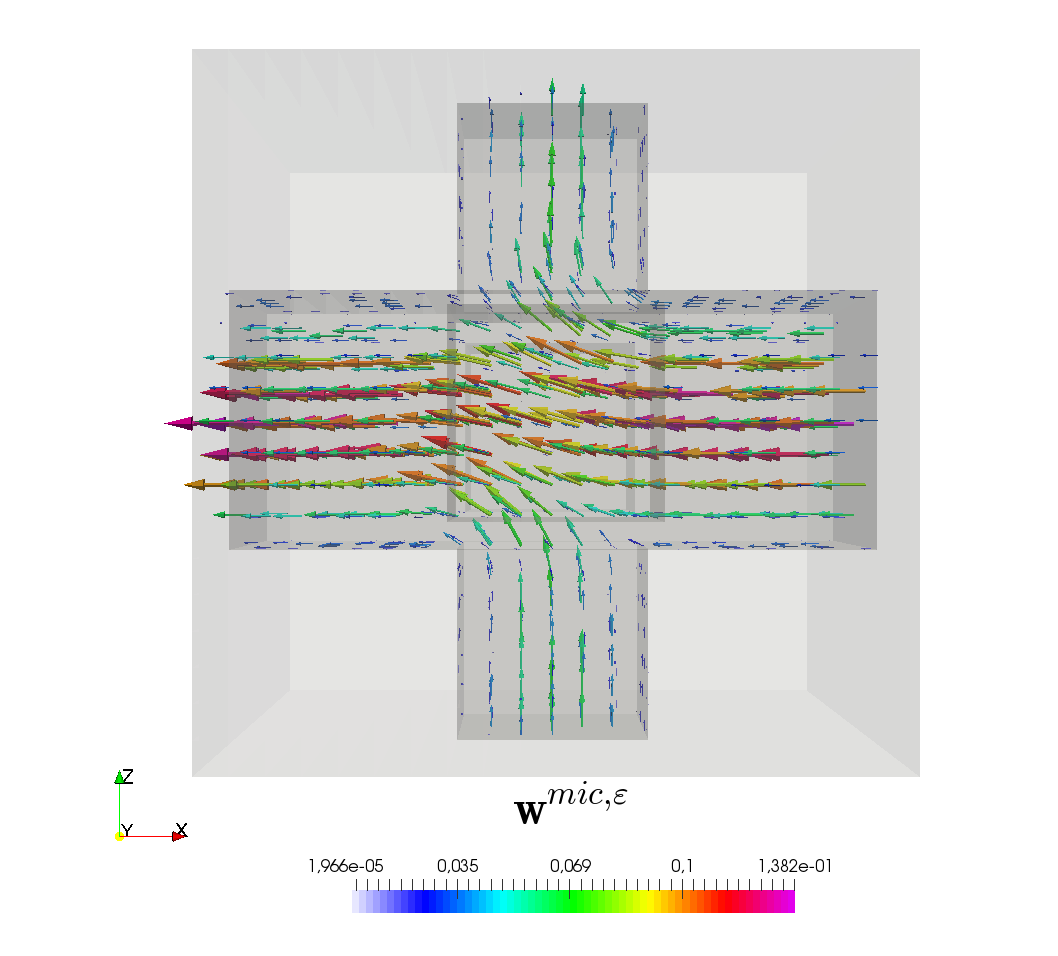}
			\caption{Reconstructed displacement $\wb^{\mic,\veps}$ in 2D view.}
		\end{subfigure}
		\begin{subfigure}{0.49\linewidth}	\centering
			\includegraphics[width= 0.95\linewidth]{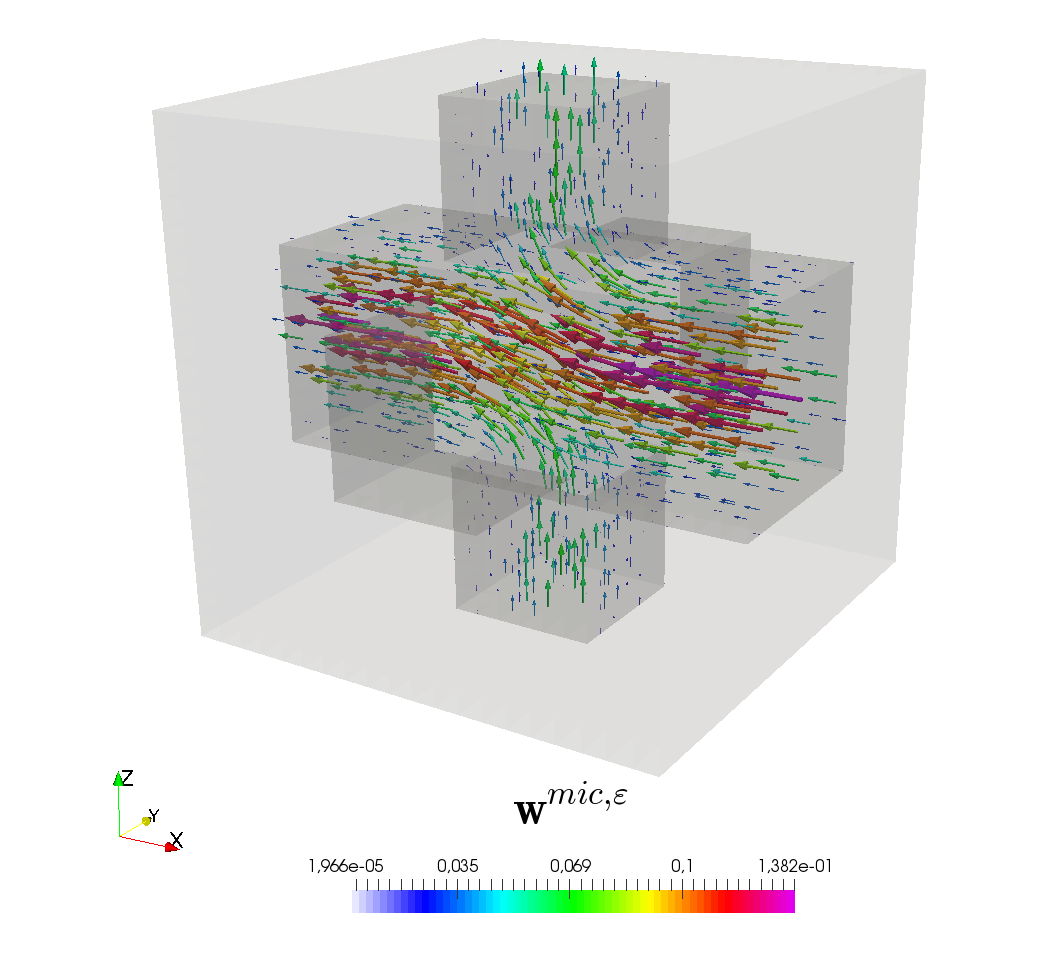}
			\caption{Reconstructed displacement $\wb^{\mic,\veps}$ in 3D view.}
		\end{subfigure}
		\caption{Reconstructed microscopic fields $\ub^{\mic,\veps}$ and $\wb^{\mic,\veps}$  on microstructure cell $Y$.}
		\label{fig_reconstructed}
	\end{figure} 
	
	For the purpose of numerical modeling, we recover the dimensionless macroscopic quantities on the microscopic cell. The macroscopic quantities were taken from the solution of BVP I, the reconstruction was made once again in \textit{SfePy}.  As the recovery cell was taken the cell in the center of the porous specimen.
	
\begin{figure}[ht]		\centering
		\begin{subfigure}{0.49\linewidth}
			\centering
			\includegraphics[width=0.95\linewidth]{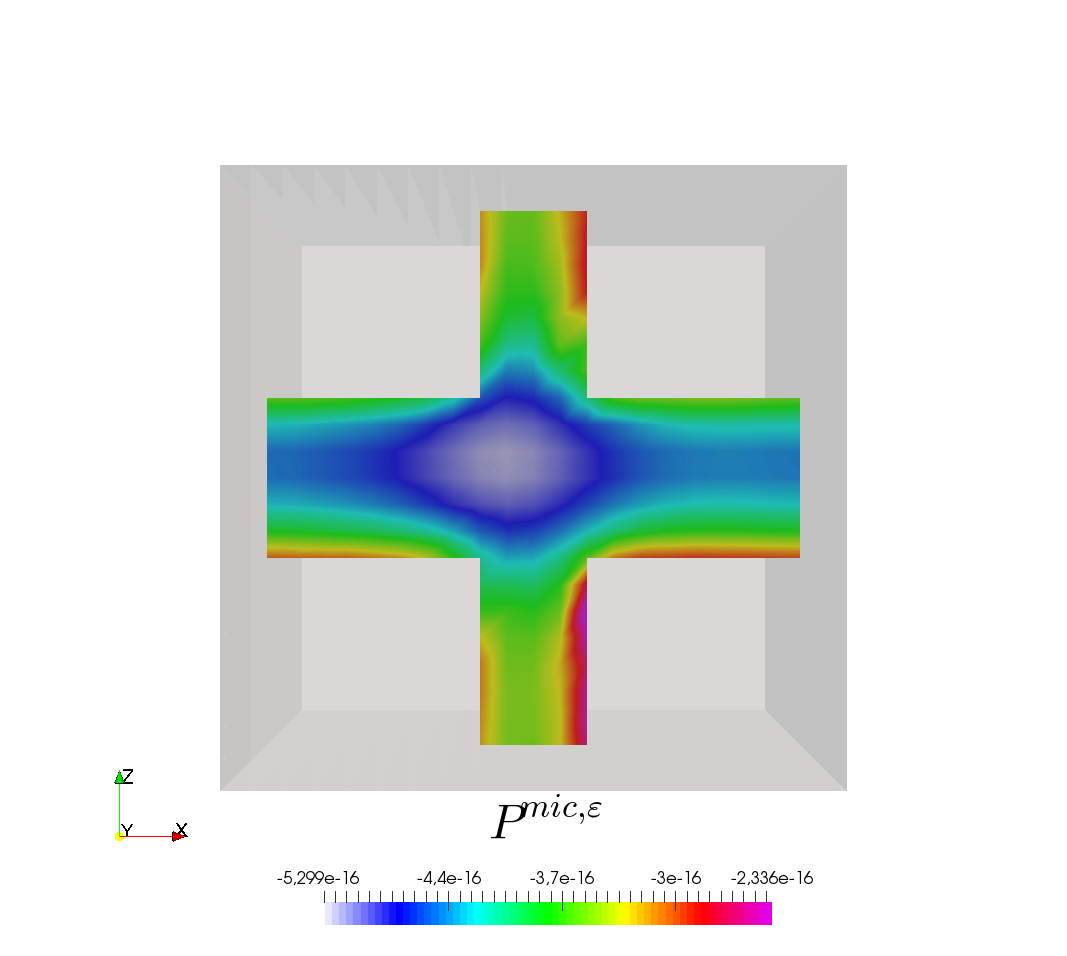}
			\caption{Reconstructed pressure $P^{\mic,\veps}$  in 2D view.}
		\end{subfigure}
		\begin{subfigure}{0.49\linewidth}	\centering
			\includegraphics[width= 0.95\linewidth]{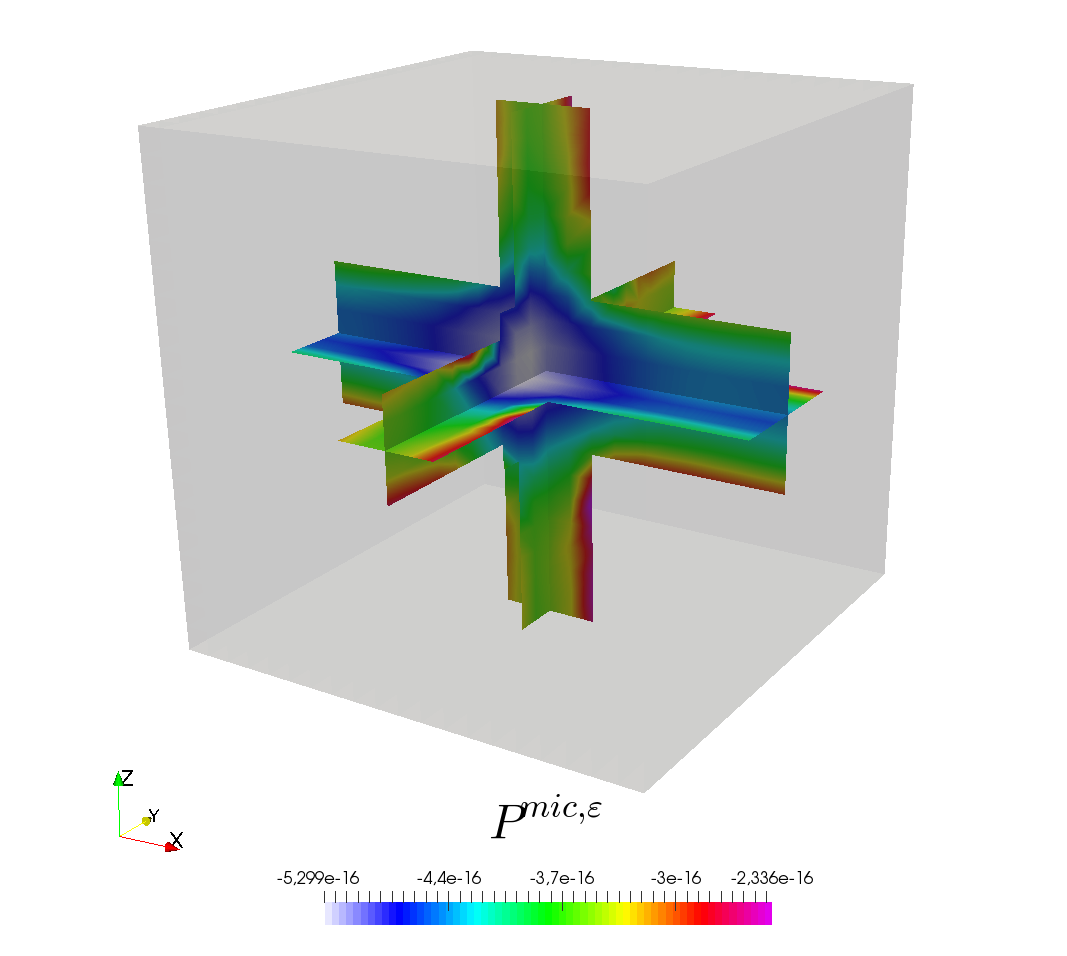}
			\caption{Reconstructed pressure $P^{\mic,\veps}$  in 3D view.}
		\end{subfigure}
		\begin{subfigure}{0.49\linewidth}
			\centering
			\includegraphics[width=0.95\linewidth]{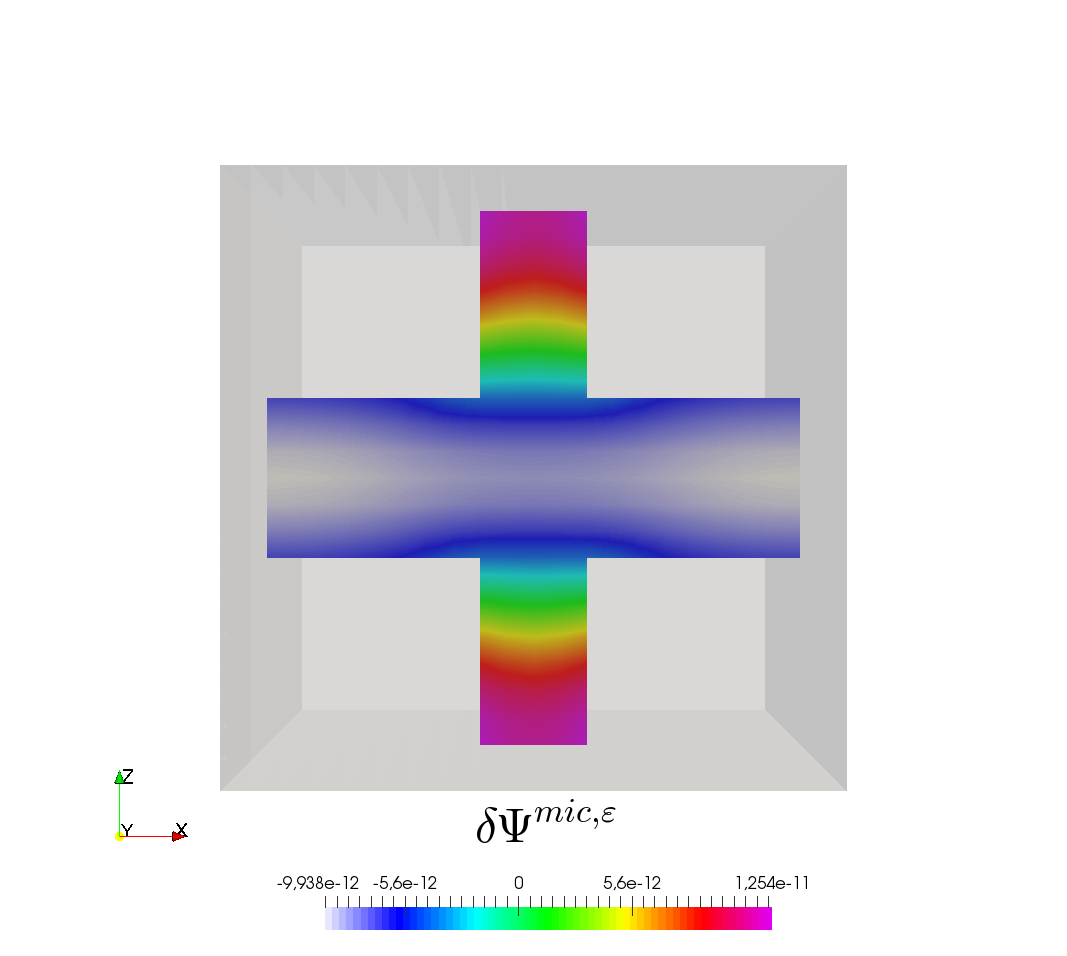}
			\caption{Reconstructed potential $\dPsi^{\mic,\veps}$ in 2D view.}
		\end{subfigure}
		\begin{subfigure}{0.49\linewidth}	\centering
			\includegraphics[width= 0.95\linewidth]{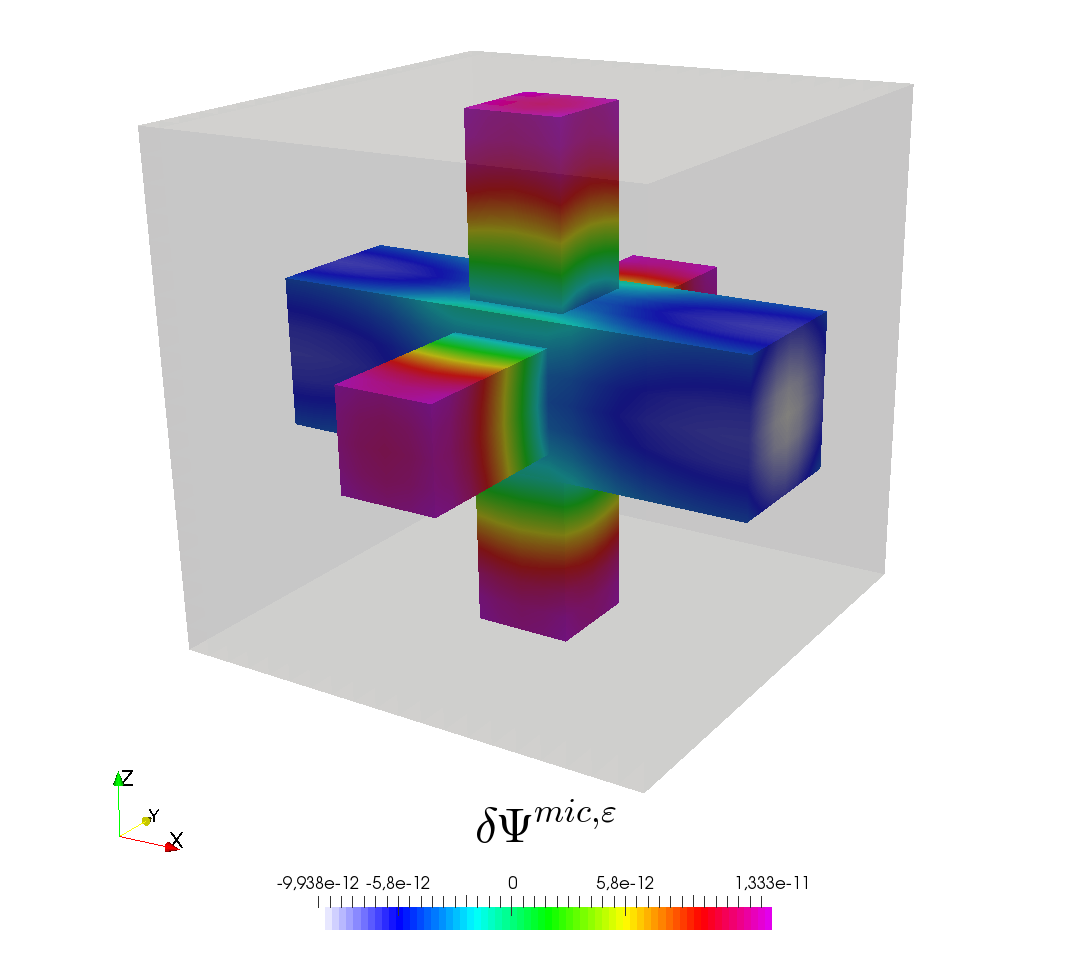}
			\caption{Reconstructed potential $\dPsi^{\mic,\veps}$ in 3D view.}
		\end{subfigure}
		\caption{Reconstructed microscopic fields $P^{\mic,\veps}$ and $\dPsi^{\mic,\veps}$ on microstructure cell $Y$.}
		\label{fig_reconstructed2}
	\end{figure}

	 As it can be seen in Fig.~\ref{fig_reconstructed}  in 2D a 3D views (depending on the specimen orientation \wrt the observer), the swelling occurs inward to the canal on the microscopic level. The reconstructed velocity field $\wb^{\mic,\veps}$ shows the electrolyte passing mainly through $y_1$- and $y_3$-directions of the canal. The reconstructed pressure $P^{\mic,\veps}$ is shown in the top half of Fig.~\ref{fig_reconstructed2}, where 3D view is illustrated using slices through all three axes of the canal. The pressure is lowest in the canal center, while highest near the solid-fluid interface. This could be traced down to the response to the solid part swelling, but is somewhat unsymmetrical due to the connections to the other macroscopic quantities. Finally,  the visualization of the reconstructed potential $\dPsi^{\mic,\veps}$ can be found in the lower half of Fig.~\ref{fig_reconstructed2}. As was mentioned earlier, the ionic potentials are locally constant. And truly, reconstructed $\Phi_\valb^{\mic,\veps}$ are constant on the whole fluid part of the microscopic cell $Y_f$ with values $\Phi_1^{\mic,\veps}=5.003\times 10^{-13}$ and $\Phi_2^{\mic,\veps}=1.543\times 10^{-12}$.

\section{Conclusion}\label{sec_conclusion}
We applied the two-scale homogenization of the deformable porous medium saturated by two-component electrolyte and implemented the resulting two-scale model in our  open source finite element code \textit{SfePy}. The upscaled model for this type of media was derived in \cite{allaire2015ion}, however, without any computational analysis, or illustrative examples. Therefore, up to our knowledge, the computational study presented in our paper provides first quantitative analysis of the considered medium. Its behavior is illustrated in terms of the example which mimics an experiment. In the computational study, dependence of the homogenized coefficients on the fluid volume fraction was presented, whereby a symmetric geometry of the reference periodic cell generating the porous medium was employed; influences of anisotropy  and other geometry-related features will be studied in our further research.

The model describes the steady state of the electrolyte flow in the solid skeleton made of an elastic electric conductor.
The homogenization is applied to the linearized model is obtained for the equilibrium reference state which is defined under the assumptions of zero fluxes (fluid and solid velocities and electroneutrality in bulk of the electrolyte). Moreover, the surface electric charge is a given constant on the solid-fluid interface. 

It is important to note that the homogenized model, thereby values of the homogenized coefficients are specific to a given microstructure size, as the result of the scale-dependent effects associated with the viscous flow and the electric double layer. Thus, a given characteristic microstructure size $\ell$ determines the Peclet number and another coefficient influencing the equilibrium electric potential.

By virtue of the ``downscaling'' procedure of the homogenization, the macroscopic fields can be used to reconstruct responses at the microscopic level. In particular, the corrector functions combined with local values of the macroscopic fields provide  two-scale functions representing the $\veps$-order fluctuating parts of the global pressure, the streaming potentials of the species, and of the displacements. In analogy, the reconstruction provides also the fluid velocity and the electric potential associated with the double layer potential which both are purely two-scale functions relevant to the microscopic cell. The perturbations of ionic concentrations from the electroneutrality state are recovered using the perturbed electric potentials.

There are further extensions of the present two-scale model and its computational implementation. To treat non-steady flows, the fluid-structure interaction will be more involved, thus, leading to a strong coupling between the flow, ionic concentrations and deformation. The scale decoupling procedure will be more complicated and will lead to fading memory effects of the macroscopic responses, as the homogenized coefficients will serve for time convolution kernels, \cf \cite{Auriault1993,rohan-etal-jmps2012-bone}.

\paragraph{Acknowledgment}
 This research is supported by part by project GACR 16-03823S of the Scientific Foundation of the Czech Republic and by the project  LO 1506 of the Czech Ministry of Education, Youth and Sports. The work was also supported from European Regional Development Fund-Project
 „Application of Modern Technologies in Medicine and Industry” (No.
 CZ.02.1.01/0.0/0.0/17\_048/0007280).
 

\appendix
\section{Unfolding homogenization}\label{sec_A}
 The unfolding homogenization method is based on the properties of unfolding operator $\Tuft$  which is similar to the dilatation operator.  
 By virtue of the coordinate decomposition into "coarse" and "fine" parts, any function $\psi=\psi(x)$ can be unfolded into a function of $x$ and $y$. The convergence results in the unfolded
domains $\Om \times Y$ can be found in  \cite{cioranescu2008periodic}. By virtue of its definition, for specific subsequences of $\veps$, domain $\Om = ]0,L[^d$ contains    the ``entire'' periods $\veps Y$, thus
\begin{equation*}\label{eq:3}
\begin{split}
\hat \Om^\veps & = \mbox{interior} \bigcup_{\zeta \in \Xi^\veps} Y_\zeta^\veps\;, \quad Y_\zeta^\veps= \veps (\ol{Y} + \zeta )\\
\mbox{ where } \Xi^\veps & = \{\zeta \in \ZZ^3\,|\; \veps (\ol{Y} + \zeta) \subset \Om\}\;.
\end{split}
\end{equation*}

For all $z \in \RR^3$,  let $[z]$ be the unique integer such that $z- [z] \in Y$.
We may write $z = [z]+\{z\}$ for all $z\in \RR^3$, so that
for all $\veps >0$, we get the unique decomposition
\begin{equation}\label{eq:3a}
x = \veps\left ( \left [\frac{x}{\veps}\right ] + \left \{\frac{x}{\veps}\right \}\right) =  \xi + \veps y
\quad \forall x \in \RR^3\;,\quad \xi = \veps\left [\frac{x}{\veps}\right ]\;.
\end{equation}
Based on this decomposition, the periodic unfolding operator
$\Tuftxt{}:  L^2(\Om;\RR) \rightarrow L^2(\Om \times Y;\RR)$ is defined as follows: for
 any function $v \in L^1(\Om;\RR)$, extended to $L^1(\RR^3;\RR)$ by zero outside $\Om$,
i.e. $v=0$ in $\RR^3 \setminus \Om$,
\begin{equation*}
\Tuf{v}(x,y) =
\left \{
\begin{array}{ll}
v\left( \veps \displaystyle  \left [\frac{x}{\veps}\right ] + \veps y \right)\;,
\quad &  x \in \hat\Om^\veps, y \in Y\;, \\
0 & \mbox{ otherwise }. \\
\end{array}
\right .
\end{equation*}
Unfolding operator $\Tuft$  has the following three important properties: For all functions $\psi$ and $\chi$:
  \begin{eqnarray}
  (i) & & \Tuf{\psi(x)\chi(x)}=\Tuf{\psi(x)}\Tuf{\chi(x)}, \\
  (ii) & & \int\limits_\Om{\psi(x)}dx = \int\limits_\Om \frac{1}{|Y|}\int\limits_Y\Tuf{\psi}(x,y)dx dy = \intY_{\Om}\Tuf{\psi}(x,y),\\ \label{uo_3}
  (iii) & & \Tuf{\nabla_x\psi(x)}=\frac{1}{\varepsilon}\nabla_y(\Tuf{\psi}(x,y)).
  \end{eqnarray}
By $\Mcal_Y(\cdot)$ we denote the average operator over $Y$, if $\Tuf{w^\veps}\rightharpoonup \hat w$ weakly in $L^p(\Om\times Y)$, then ${w^\veps}\rightharpoonup \Mcal_Y(\hat w)$
weakly in $L^p(\Om)$.
For any $D\subset Y$,
$\intYsmall_{D} = \frac{1}{|Y|}\int_{D}$; the analogical notation is employed for any $A\subset Z$, thus $\intYsmall_{A} = \frac{1}{|Z|}\int_{A}$. Further, for any $D\subset Y$, $\Hpdb(D)$ is the Sobolev space $\Wb^{1,2}(Y) = \Hdb(Y)$ of vector-valued Y-periodic functions (indicated by the subscript $\#$).

   \par Importantly, the unfolding operator also transforms the integration in domain $\Om$ to $\Om\times Y$, so that standard means of the weak convergence in Lebesgue spaces $L^q(\Om\times Y)$ can be employed. For more details see  \cite{cioranescu2008periodic}.

 \section{Introduction of parameter $\veps$ into the dimensionless system}\label{sec_B}
 Here we clarify, how the scale parameter $\veps$ (see Section~\ref{sec:dimless}) is introduced into the system of equations \Eq{eq_balance}-\Eq{eq_incompressibility} along with the derivation of the dimensionless form \Eq{eq_adim_start}-\Eq{eq_deform_adim_stop}. 
 \paragraph{Modified Stokes problem}
The characteristic pressure $p_c$ is expressed using the ideal gas law, 
 \begin{equation}\label{eq_pc}
 p_c=c_ck_BT.
 \end{equation}
 Then, by inserting the dimensionless quantities \Eq{dimless_q} and the dimensionless operator \Eq{dimless_o} into \Eq{eq_stokes},  we get
 \begin{equation}\label{eq_app_B_2}
  \nabla\pe-\frac{v_c\eta_f}{L_cp_c}\Delta\wbe=\frac{L_c}{p_c}\fb -\frac{ec_c\Psi_c}{p_c}\sum\limits_{\valblim}^{2}\zj\cje\nabla\Psie\;,
 \end{equation}
 hence \Eq{dimless_f} introduces the dimensionless force $\fbh=\frac{L_c}{p_c}\fb$. Upon  substituting  expressions $\Psi_c=k_B T/e$ and \Eq{eq_pc} into \Eq{eq_app_B_2}, we get $\frac{ec_c\Psi_c}{p_c}=1$. Further, according to
\cite{lemaire2011multiscale},  the ration between the velocity and pressure magnitudes
 $\lambda_c:=\frac{v_c\eta_f}{L_cp_c}$ is obtained by the dimensional analysis of the Darcy law which can also represent the viscous flow in pores. This yields 
 \begin{equation*}\label{darcy}
 v_c=\frac{kp_c}{\eta_fL_c},
 \end{equation*}
 where $k$ denotes the intrinsic permeability (units [m$^2$]) depending only on the size of the micropores,  $k\sim l^2$, hence holds and
 \begin{equation}\label{eq-B3}
 \lambda_c=\frac{v_c\eta_f}{L_c p_c}=\frac{k}{L_c^2 }\sim\frac{l^2}{L_c^2 }=\veps^2.
 \end{equation}
 Consequently from \Eq{eq_pc}-\Eq{eq-B3}, the dimensionless form \Eq{eq_stokes} reads, 
 \begin{equation}\label{eq-B4}
 \nabla\pe-\veps^2\Delta\wbe=\fb^* -\sum\limits_{\valblim}^{2}\zj\cje\nabla\Psie.
 \end{equation}
 
  \paragraph{Electrostatics}
  Upon substituting \Eq{dimless_q} and \Eq{dimless_o} in the Gauss-Poisson equation \Eq{eq_elektrokinetic}, it yields
  \begin{equation}\label{eq_dimless_electrokinetics1}
  \frac{\mathcal{E}\Psi_c}{L_c^2}\Delta\Psie =-ec_c\sum\limits_{\valblim}^{2}\zj\cje.
  \end{equation}
  Using the Debye length definition \Eq{debye} and parameter $\betapar=l^2(\lambda_D\sum_{\valblim}^{2}\zj^2)^{-1}$, we may express the characteristic concentration $c_c$ as
  \begin{equation}
  c_c=\frac{\mathcal{E}k_b T}{(e\lambda_D)^2\sum_{\valblim}^{2}\zj^2}=\betapar\frac{\mathcal{E}k_b T}{(el)^2}.
  \end{equation}
  By substituting $c_c$ and $\Psi_c$ into \Eq{eq_dimless_electrokinetics1}, we get
  \begin{equation}\label{eq_dimless_electrokinetics2}
  \frac{\mathcal{E}k_b T}{eL_c^2}\Delta\Psie =-\frac{e\betapar\mathcal{E}k_b T}{(el)^2 }\sum\limits_{\valblim}^{2}\zj\cje,
  \end{equation}
  so that $c_cL_c = \veps^2$, hence \Eq{eq_dimless_electrokinetics1} reads
  \begin{equation}\label{eq_dimless_electrokinetics}
  \veps^2\Delta\Psie =-\betapar\sum\limits_{\valblim}^{2}\zj\cje.
  \end{equation}
  Similarly, by inserting \Eq{dimless_o}-\Eq{dimless_f} into \Eq{eq_elektrokinetic_bc}, we get
  \begin{equation}\label{eq_dimless_elkin_bc1}
  \frac{\mathcal{E}\Psi_c}{L_c}\nabla\Psie\cdot\nb=-\Sigma_c\Sigma^*.
  \end{equation}
 After a few easy adjustments we get its dimensionless form as follows
  \begin{eqnarray}\label{eq_dimless_elkin_bc}
  \veps\nabla\Psie\cdot\nb=-\frac{el\Sigma_c}{\mathcal{E}k_BT}\Sigma^* = -\Nsig\Sigma^*\onome,
  \end{eqnarray}
  where $\Nsig=\frac{el\Sigma_c}{\mathcal{E}k_BT}$ is the ratio between electrical and thermal energy and it is usually of order $\mathcal{O}(1)$ in $\veps$, \cite{moyne2002electro}.

\bibliographystyle{unsrt}

\begin{thebibliography}{10}
\bibitem[Allaire et al. (2010)]{allaire2010homogenizationA}
 Allaire, G., Mikeli{\'c}, A., Piatnitski, A.,
 \textit{Homogenization of the linearized ionic transport equations in rigid periodic porous media}, Journal of Mathematical Physics \textbf{51}(12), 123103 (2010).


\bibitem[Allaire et al. (2013)a]{allaire2013asymptotic}
 Allaire, G., Dufr{\^e}che, J.-F., Mikeli{\'c}, A., Piatnitski, A.,
\textit{Asymptotic analysis of the Poisson--Boltzmann equation describing electrokinetics in porous media}, Nonlinearity \textbf{26}(3), (2013).

\bibitem[Allaire et al. (2013)b]{allaire2013ion}
Allaire, G., Brizzi, R., Dufr{\^e}che, J.-F., Mikeli{\'c}, A., Piatnitski, A.,
\textit{Ion transport in porous media: derivation of the macroscopic equations using upscaling and properties of the effective coefficients}, Computational Geosciences \textbf{17}(3), (2013).

\bibitem[Allaire et al. (2015)]{allaire2015ion}
Allaire, G., Bernard, O., Dufr{\^e}che, J.-F., Mikeli{\'c}, A., 
\textit{Ion transport through deformable porous media: derivation of the macroscopic equations using upscaling}, Computational and Applied Mathematics, (2015).

\bibitem[Amirat and Shelukhin (2008)]{amirat2008electroosmosis}
Amirat, Y., Shelukhin, V., 
\textit{Electroosmosis law via homogenization of electrolyte flow equations in porous media}, Journal of Mathematical Analysis and Applications \textbf{342}(2), (2008).

\bibitem[Andreasen and Sigmund (2013)]{andreasen2013topology}
Andreasen, C. S., and Sigmund, O., 
\textit{Topology optimization of fluid--structure-interaction problems in poroelasticity}, Computer Methods in Applied Mechanics and Engineering \textbf{258}, (2013).

\bibitem{Auriault1993}
J.L. Auriault and C.~Boutin.
\newblock Deformable porous media with double porosity. quasi-statics. ii:
  Memory effects.
\textit{Transport in porous media}, \textbf{10}(2):153--169, (1993).


\bibitem[Cimrman (2014)]{sfepy}
Cimrman, R., 
\textit{SfePy-write your own FE application}, arXiv preprint arXiv:1404.6391, (2014).


\bibitem[Cioranescu et al. (2008)]{cioranescu2008periodic}
Cioranescu, D., Damlamian, A., Griso, G., 
\textit{The periodic unfolding method in homogenization}, SIAM Journal on Mathematical Analysis \textbf{40}(4), (2008).

\bibitem[Frank et al. (2011)]{frank2011numerical}
Frank, F., Ray, N., Knabner, P., 
\textit{Numerical investigation of homogenized Stokes--Nernst--Planck--Poisson systems}, Computing and visualization in science \textbf{14}(8), (2011).

\bibitem[Hunter (2001)]{hunter2001foundations}
Hunter, R. J., 
\textit{Foundations of colloid science}, Oxford University Press, (2001).

\bibitem[Jones et al. (2001)]{scipy}
Jones, E., Oliphant, T., Peterson, P. and others
\textit{{SciPy}: Open source scientific tools for {Python}}, "http://www.scipy.org/", (2001--), [Online; accessed <today>].

\bibitem[Lemaire et al. (2006)]{lemaire2006multiscale}
Lemaire, T., Na{\"\i}li, S., R{\'e}mond, A.,
\textit{Multiscale analysis of the coupled effects governing the movement of interstitial fluid in cortical bone}, Biomechanics and modeling in mechanobiology \textbf{5}(1), (2006).

\bibitem[Lemaire et al. (2010)a]{lemaire2010modelling}
Lemaire, T., Kaiser, J., Na{\"\i}li, S., Sansalone, V.,
\textit{Modelling of the transport in electrically charged porous media including ionic exchanges}, Mechanics Research Communications \textbf{37}(5), (2010).

\bibitem[Lemaire et al. (2010)b]{lemaire2010multiphysical}
Lemaire, T., Na{\"\i}li, S., Sansalone, V.,
\textit{Multiphysical modelling of fluid transport through osteo-articular media}, Anais da Academia Brasileira de Ci{\^e}ncias \textbf{82}(1), (2010).

\bibitem{lemaire2011multiscale}
Lemaire, T., Capiez-Lernout, E., Kaiser, J., Na{\"\i}li, S., Sansalone, V., Rohan, E.
	\newblock {A multiscale theoretical investigation of electric measurements in living bone},
	\textit{Bulletin of mathematical biology},
	\textbf{73}(11):{2649--2677}, (2011).



\bibitem[Looker and Carnie (2006)]{looker2006homogenization}
Looker, J. R., Carnie, S. L.,
\textit{Homogenization of the ionic transport equations in periodic porous media}, Transport in porous media \textbf{65}(1), (2006).

\bibitem[Mikeli{\'c} and Wheeler (2012)]{mikelic2012interface}
Mikeli{\'c}, A., Wheeler, M. F.,
\textit{On the interface law between a deformable porous medium containing a viscous fluid and an elastic body}, Mathematical Models and Methods in Applied Sciences \textbf{22}(11), (2012).

\bibitem[Moyne and Murad (2002)]{moyne2002electro}
Moyne, C., Murad, M. A.,
\textit{Electro-chemo-mechanical couplings in swelling clays derived from a micro/macro-homogenization procedure}, International Journal of Solids and Structures \textbf{39}(25), (2002).

\bibitem[Moyne and Murad (2006)]{moyne2006two}
Moyne, C., Murad, M. A.,
\textit{A two-scale model for coupled electro-chemo-mechanical phenomena and Onsager’s reciprocity relations in expansive clays: I homogenization analysis}, Transport in Porous Media \textbf{62}(3), (2006).

\bibitem[Nguyen et al. (2009)]{nguyen2009numerical}
Nguyen, V.-H., Lemaire, T., Naili, S.,
\textit{Numerical study of deformation-induced fluid flows in periodic osteonal matrix under harmonic axial loading}, Comptes Rendus Mecanique \textbf{337}(5), (2009).

\bibitem[O'Brien and White (1978)]{obrien1978electrophoretic}
O'Brien, R. W., White, L. R.,
\textit{Electrophoretic mobility of a spherical colloidal particle}, Journal of the Chemical Society, Faraday Transactions 2: Molecular and Chemical Physics \textbf{74}, (1978).

\bibitem[Ray et al. (2012)a]{ray2012multiscale}
Ray, N., van Noorden, T., Frank, F., Knabner, P.,
\textit{Multiscale modeling of colloid and fluid dynamics in porous media including an evolving microstructure}, Transport in porous media \textbf{95}(3), (2012).

\bibitem[Ray et al. (2012)b]{ray2012rigorous}
Ray, N., Muntean, A., Knabner, P.,
\textit{Rigorous homogenization of a stokes--nernst--planck--poisson system}, Journal of Mathematical Analysis and Applications \textbf{390}(1), (2012).

\bibitem{rohan-etal-jmps2012-bone}
E.~Rohan, S.~Naili, R.~Cimrman, and T.~Lemaire.
\newblock Multiscale modeling of a fluid saturated medium with double porosity:
  Relevance to the compact bone.
\textit{ Journal of the Mechanics and Physics of Solids}, \textbf{60}(5):857--881,
  (2012).

\bibitem{Rohan-AMC}
E.~Rohan and V.~Lukeš.
\newblock Modeling nonlinear phenomena in deforming fluid-saturated porous
  media using homogenization and sensitivity analysis concepts.
  \textit{ Applied Mathematics and Computation}, 267:583–595, 2015.

\bibitem{rohan-etal-CMAT2015}
E.~Rohan, S.~Naili, and T.~Lemaire.
\newblock Double porosity in fluid-saturated elastic media: deriving effective
  parameters by hierarchical homogenization of static problem.
\newblock {\textit{ Continuum Mechanics and Thermodynamics}}, \textbf{28}(5):1263–1293, (2015).  
  
\bibitem[Rohan et al. (2017)]{rohan2017darcy}
Rohan, E., Turjanicov{\'a}, J., Luke{\v{s}}, V.,
\textit{A Darcy-Brinkman model of flow in double porous media--Two-level homogenization and computational modelling}, Computers \& Structures, (2017).

\bibitem{Sandstrom-Larssen-CMAME2016}
C.~Sandström, F.~Larsson, and K.~Runesson.
\newblock Homogenization of coupled flow and deformation in a porous material.
\textit{ Computer Methods in Applied Mechanics and Engineering},
\textbf{308}:535–551, (2016).

\bibitem{Schmuck-Bazant-SIAM2015}
Schmuck,M., Bazant, M. Z.
\newblock Homogenization of the Poisson-Nernst-Planck equations for ion transport in charged porous media. \textit{SIAM Journal on Applied Mathematic}, \textbf{75}(5):1369–1401, (2015).  

\end{thebibliography}

\end{document}